\documentclass[12pt]{article}
\usepackage{geometry}
\usepackage[cp1252]{inputenc}
\usepackage[T1]{fontenc}
\usepackage{lmodern}
\usepackage{graphicx}
\usepackage{amsmath}
\usepackage{amssymb}
\usepackage{amsthm}
\usepackage{amsfonts}
\usepackage{authblk}
\usepackage{epstopdf}
\usepackage{verbatim}
\usepackage{multicol}
\usepackage{tabularx}
\usepackage{natbib}
\usepackage{graphicx}
\usepackage{multirow}
\usepackage{array}
\usepackage{color}
\usepackage{float}
\restylefloat{table}
\usepackage{xcolor}
\usepackage{lscape}
\usepackage{ragged2e}
\usepackage{booktabs}
\usepackage{topcapt}
\usepackage{rotating}
\usepackage{setspace}
\usepackage{times}
\usepackage{txfonts}
\usepackage{bm}
\usepackage{titlesec}
\usepackage{indentfirst}
\usepackage{enumitem}
\usepackage{algorithm,algorithmic}

\newcolumntype{Y}{>{\raggedleft\arraybackslash}X}
\newcolumntype{C}{>{\centering\arraybackslash}X}
\newcommand{\citeS}[1]{\citeauthor{#1}'s (\citeyear{#1})}
\allowdisplaybreaks
\date{This version: \today \\[0.1ex] First draft: May 22, 2019}

\author[1]{Jean-Fran\c cois B\'egin\footnote{Corresponding author.}$^{\!,}$}
\author[2]{Mathieu Boudreault}
\affil[1]{Simon Fraser University}
\affil[2]{Université du Québec à Montréal}

\title{Likelihood Evaluation of Jump-Diffusion Models Using Deterministic Nonlinear Filters\footnote{For their useful comments, the authors would like to thank 
Maciej Augustyniak, 
Genevi\`eve Gauthier, 
Jonathan Grégoire, 
Fr\'ed\'eric Godin, 
as well as the seminar participants at the 2019 Quantact Workshop on Financial Mathematics and the 23rd International Congress on Insurance: Mathematics and Economics. Bégin wishes to acknowledge the financial support of the National Science and Engineering Research Council of Canada (NSERC), Simon Fraser University and the NVIDIA Corporation. Boudreault would like to thank the financial support of NSERC.}}

\titleformat{\section}{\normalfont\bfseries}{\thesection}{1em}{}
\titleformat{\subsection}{\normalfont\bfseries}{\thesubsection}{1em}{}
\titleformat{\subsubsection}{\normalfont\bfseries}{\thesubsubsection}{1em}{}

\titlespacing*{\section}{0pt}{0pt}{0pt}
\titlespacing*{\subsection}{0pt}{0pt}{0pt}
\titlespacing*{\subsubsection}{0pt}{0pt}{0pt}

\begin{document}
\makeatletter
\g@addto@macro{\normalsize}{%
    \setlength{\abovedisplayskip}{4pt}
    \setlength{\abovedisplayshortskip}{4pt}
    \setlength{\belowdisplayskip}{4pt}
    \setlength{\belowdisplayshortskip}{4pt}}
\makeatother

\maketitle
\thispagestyle{empty}

\begin{abstract}
In this study, we develop a deterministic nonlinear filtering algorithm based on a high-dimensional version of Kitagawa (1987) to evaluate the likelihood function of models that allow for stochastic volatility and jumps whose arrival intensity is also stochastic. We show numerically that the deterministic filtering method is precise and much faster than the particle filter, in addition to yielding a smooth function over the parameter space. We then find the maximum likelihood estimates of various models that include stochastic volatility, jumps in the returns and variance, and also stochastic jump arrival intensity with the S\&P 500 daily returns. During the Great Recession, the jump arrival intensity increases significantly and contributes to the clustering of volatility and negative returns.
\vspace{0.25cm}

\end{abstract}
\vspace{-0.25cm}
\newpage
\setcounter{page}{1}

\section{Introduction and Motivation}


Over the last thirty years, jump-diffusion models (with stochastic volatility) have become increasingly popular for their ability to replicate important stylized facts such as heavy tails, no autocorrelation in the returns, volatility clustering, etc. \citep{cont2007volatility}. Yet, parameter estimation of these models is cumbersome as stochastic volatility and jumps are latent---or not directly observed. Therefore, the first studies that looked into that problem focused mostly on stochastic volatility models using the (generalized) method of moments, quasi-, simulated and approximate maximum likelihood, and Bayesian methods.\footnote{The first study looking into parameter estimation of stochastic volatility models is \citet{taylor2008modelling} which was based upon the method of moments. Generalizations of the method of moments were proposed by \citet{melino1990pricing}, \citet{andersen1996gmm}, \citet{pan2002jump}. Quasi-maximum likelihood approaches were suggested by \citet{nelson1988time} and \citet{harvey1994multivariate}; \citet{danielsson1994stochastic} and \citet{brandt2002simulated} used simulated maximum likelihood whereas approximate maximum likelihood were introduced by \citet{ait2002maximum}, \citet{bates2006maximum} and \citet{ai2007maximum}. Finally, another family of approaches based upon the Bayesian paradigm was used in this literature, namely \citet{shephard1993fitting}, \citet{jacquier1994bayesian}, \citet{johannes1999state} and \citet{eraker2001mcmc}.}

Most recent progress in parameter estimation of jump-diffusion models has been accomplished with numerical recursive prediction-update algorithms, namely sequential Monte Carlo (SMC) methods and discrete nonlinear filters (DNF). Such methods provide the posterior distribution of latent state variables conditional on current and past observations in a recursive manner.\footnote{A classic example of prediction-update algorithms is the \citet{kalman1960new} filter which provides the exact posterior distribution of the state variables in a Gaussian and linear framework.} As a by-product of the method, one can obtain the likelihood function. On the one hand, sequential Monte Carlo (SMC) methods---so-called particle filters---as introduced by \citet{gordon1993novel}, have been used to compute and maximize the likelihood function \citep{johannes2009optimal, christoffersen2010volatility, pitt2014simulated, begin2019idiosyncratic, bardgett2019inferring}. Although the SMC method is very flexible, it is Monte Carlo-based and thus computationally intensive. Moreover, the likelihood function is random and not smooth for a finite set of particles, which can be a significant issue for frequentist-based inference.\footnote{\citet{malik2011particle} introduced a resampling algorithm that smoothes the likelihood function but according to \citet{creal2012survey}, this only works when the state dimension is one, which is typically the case when only stochastic volatility is included whereas jumps are excluded.}

On the other hand, one can approximate the prediction density as well as the likelihood function with numerical integration schemes; this is the framework put forward by  \citet{kitagawa1987non}. As such, the dimension of the integration problem increases with the number of state variables which explains why many authors solely focused on stochastic volatility models \citep{fridman1998maximum, watanabe1999non, bartolucci2001maximum, clements2006estimating, langrock2012some} whereas jumps were ignored.

In this article, we present a DNF-based method for the likelihood evaluation of stochastic volatility models that include return jumps, variance jumps, as well as stochastic jump arrival intensity. Those include well-known models such as the stochastic volatility (SV) model, stochastic volatility with return jumps (SVYJ) model, and the stochastic volatility with return and variance correlated jumps (SVCJ) model, which are nested cases of the general model. We then assess the accuracy of the proposed methodology and compare it with the most common SMC method---the sequential importance resampling (SIR) of \citet{gordon1993novel}. Finally, we apply the method to a typical financial time series: the S\&P 500 index. 

Our main contributions and findings are as follows. We first develop a prediction-update algorithm for the SVCJSI model---and its nested models. The proposed approach reduces the dimensionality of the resulting integrals when compared to a naive application of \citet{kitagawa1987non}. As a result, the proposed DNF method is both accurate and faster than the SIR. And because the DNF yields a smooth likelihood function in the parameter space, it is ideal for numerical maximization. We thus compute the maximum likelihood estimates of the SV, SVYJ, SVCJ and SVCJSI\footnote{SVCJ with stochastic jump arrival intensity} models using S\&P 500 daily returns. We find that during the Great Recession, the jump arrival intensity increases significantly and contributes to the clustering of volatility and negative returns.

The article is organized as follows. Section~\ref{sec: models} presents the econometric framework. The DNF is explained in Section~\ref{sec:dnf}. Section~\ref{sec:accuracy} discusses the accuracy and the reliability of the DNF method. The speed-accuracy trade-off is assessed in Section~\ref{sec:speed}. An empirical application is presented in Section~\ref{sec:empirical}. Finally, some concluding remarks are provided in Section~\ref{sec:conclusion}.

\section{Framework}\label{sec: models}


This section describes the continuous- and discrete-time frameworks. We then specifically introduce the four models that are used throughout this paper, namely the SV, SVYJ, SVCJ, and the full model with stochastic jump arrival intensity, known as SVCJSI.

\subsection{Continuous-Time Framework}

We fix a filtered probability space $(\Omega,\mathcal{F},\mathbb{F},\mathbb{P})$ and a filtration $\mathbb{F} = \{\mathcal{F}_t \, : \, t \geq 0 \}$ satisfying the usual conditions. Let $S_t$ be the time-$t$ price of a security, $V_t$ the time-$t$ instantaneous variance and $\Lambda_t$ the time-$t$ jump intensity. Under the objective measure $\mathbb{P}$, their dynamics are given by the following equations:
\begin{align}
\frac{dS_t}{S_{t^-}} = &\, \left( \mu - \overline{\alpha} \Lambda_t  \right)\, dt + \sqrt{V_{t^{-}}} \, dW_{t}^S + d \left( \sum_{n=1}^{N_t} \left( e^{Z_n^{S}} - 1 \right) \right), \notag \\[-0.2ex]
dV_t = &\, \kappa \left( \theta - V_{t^-} \right) \, dt + \sigma \sqrt{V_{t^-}} \, dW_t^{V} +d \left( \sum_{n=1}^{N_t} Z_n^{V} \right),  \notag \\[-0.6ex]
d\Lambda_t = &\, \chi \left( \omega - \Lambda_{t} \right) \, dt \hspace{0.04cm} + \xi \sqrt{\Lambda_{t}} \, dW_t^{\Lambda} , \notag
\end{align}
where $W^S$, $W^V$ and $W^{\Lambda}$ are three Brownian motions such that $d \langle W^S, W^V \rangle_t = \rho_v \, dt$, $d \langle W^S, W^{\Lambda} \rangle_t = \rho_{\lambda} \, dt$ and $d \langle W^V, W^{\Lambda} \rangle_t = 0 \, dt$. Moreover, $\{N_t\}_{\{t\geq 0\}}$ is a Cox process (doubly stochastic Poisson process) with jump arrival intensity $\Lambda_t$ and jump sizes to be further discussed. Note that only $S_t$ is observed whereas $V_t$ and $\Lambda_t$ are unobserved or latent.

The first stochastic differential equation (SDE) resembles that proposed by \citet{merton1976option} as it includes both a jump process as well as a diffusive component. Yet, in opposition to Merton, our diffusive component allows for stochastic volatility---modelled by the second SDE. The stochastic variance SDE is given by the square-root process---similar to that used by \citet{heston1993}---to which variance jumps are added. Parameter $\kappa$ denotes the speed of mean reversion, $\theta$ is the unconditional mean variance, and $\sigma$ is the so-called variance of the variance parameter. Finally, the last SDE models the stochastic jump arrival intensity; parameter $\chi$ is related to the speed of mean reversion, $\omega$ is the long run level of the stochastic jump intensity and $\xi$ is the variance of the intensity parameter.

We assume that each of the return jumps $\{Z_n^S\}_{\{n \in \mathbb{N} \}}$ are normally distributed with a mean of $\alpha + \rho_z Z_n^V$ and a standard deviation of $\delta$, i.e., $Z_n^S \sim \mathcal{N}(\alpha+\rho_z Z_n^V ,\delta^2)$
and that the variance jumps $\{Z_n^V\}_{\{n \in \mathbb{N} \}}$ are exponentially distributed with a mean of $\nu$, i.e., $Z_n^V \sim \text{Exp}(\nu)$. Similar return jump innovations have been used by \citet{bates1996jumps}, \citet{bakshi1997empirical}, \citet{duffie2000transform}, \citet{pan2002jump}, \citet{eraker2003impact}, and \citet{johannes2009optimal} in continuous-time models whereas \citet{maheu2004news}, \citet{christoffersen2012dynamic} and \citet{ornthanalai2014levy} (among others) have introduced normally distributed jumps in discrete-time models. Evidence for the presence of positive jumps in the variance process has been found by numerous authors, namely \citet{bates2000post}, \citet{duffie2000transform}, \citet{pan2002jump}, \citet{eraker2003impact} and \citet{todorov2011volatility}.

Finally, the term $\Lambda_t \overline{\alpha}$ is the jump compensator that makes sure that $\mathbb{E}^{\mathbb{P}}\left[ S_T \, \middle|\, \mathcal{F}_t \right] = S_t e^{\,\mu(T-t)}$, with $\mu$ being the annual expected rate of return on the stock in our setting. We thus have $\overline{\alpha} = {\exp\left( \alpha + \frac{1}{2}\delta^2 \right)}/{\left(1-\rho_z \nu\right)} - 1$.

\subsection{Discrete-Time Framework}
As we typically observe most continuous-time processes only in discrete-time, we thus need to discretize our continuous-time dynamics. Setting $h$ as the length of a time interval (say e.g. a month, a week or a calendar/trading day), then we let $y_t \equiv \log(S_{th}/S_{(t-1)h})$ be the time-$t$ (continuously compounded ex-dividend) return of a security. As such, the information set available to the econometrician $\mathbb{G} = \{\mathcal{G}_t \, : \, t \in \mathbb{N} \}$ is coarser than that of the model, i.e., $\mathcal{G}_t \subseteq \mathcal{F}_t$ since $\mathcal{G}_t = \sigma\left( \{y_s\}_{s=1}^t \right)$. Similarly, we define $v_t$ and $\lambda_t$ as the time-$th$ discretized variance and jump intensity processes, respectively. Again, as in the continuous-time version of the model, $v_t$ and $\lambda_t$ are latent.

The Euler-Maruyama discretization scheme applied to the continuous-time version of the model yields
\begin{align}
y_t = &\, \left( \mu - \frac{1}{2} v_{t-1} - \overline{\alpha} \lambda_{t-1} \right) h + \sqrt{v_{t-1} \, h} \varepsilon_t^{y} + {\sum_{i=1}^{n_t} z^y_{t,i}}, \notag \\
v_{t} = &\, v_{t-1} + \kappa \left( \theta - v_{t-1} \right) h + \sigma \sqrt{v_{t-1} \, h} \, \varepsilon_t^v + \sum_{i=1}^{n_t} z^v_{t,i} , \notag \\
\lambda_{t} = &\, \lambda_{t-1} + \chi \left( \omega - \lambda_{t-1} \right) h + \xi \sqrt{\lambda_{t-1} \, h} \, \varepsilon_t^{\lambda} \notag ,
\end{align}
where $n_t = N_{th} - N_{(t-1)h}$ is given by a Poisson random variable with parameter $\lambda_{t-1} h$. Moreover, the standardized innovations are normally distributed, i.e., $\varepsilon_t^{i} \sim \mathcal{N}(0,1)$ for $i \in \{\perp,v,\lambda\}$ with
\begin{equation}
\varepsilon_t^y = \sqrt{1- \rho_v^2 - \rho_{\lambda}^2} \varepsilon^{\perp}_t + \rho_v \varepsilon_t^v + \rho_{\lambda} \varepsilon_t^{\lambda} \notag
\end{equation}
obtained from the Cholesky decomposition. Finally, return jumps are also normally distributed, i.e. $z^y_{t,i} \sim \mathcal{N}\left( \alpha + \rho z_{t,i}^v, \, \delta^2 \right)$, whereas the distribution of variance jumps is still exponential, i.e., $z^v_{t,i} \sim \text{Exp}(\nu)$.

\subsection{Nested Models}

This general specification embeds numerous stochastic volatility-type models. The four specific models considered in this study are detailed below:
\begin{itemize}
\itemsep0em
\item[--] \textbf{SV}: Stochastic volatility model with no jumps, obtained by letting $\chi = \omega = \xi = 0$. It is similar to the model proposed by \cite{heston1993} in the sense that it incorporates a leverage effect. Specifications alike were also investigated by \citet{melino1990pricing}, \citet{danielsson1993accelerated}, \citet{shephard1993fitting}, and \citet{jacquier1994bayesian}, among others.
\item[--] \textbf{SVYJ}: Stochastic volatility model with jumps in return only, obtained by letting $\chi = 1$ and $\xi = \nu = 0$. It resembles the \citeS{bates1996jumps} model. In econometrics, \cite{pitt2014simulated} proposed the use of jumps in the return's dynamics.
\item[--] \textbf{SVCJ}: Stochastic volatility with simultaneous and correlated jumps in return
and variance, obtained by letting $\chi = 1$ and $\xi = 0$. The specification was first used by \citet{duffie2000transform}.
\item[--] \textbf{SVCJSI}: Stochastic volatility with simultaneous and correlated jumps in return
and variance along with stochastic jump arrival intensity. This is the most complicated and general specification used in this study.
\end{itemize}

\section{Discrete Nonlinear Filtering\label{sec:dnf}}

Because the volatility, jumps and the jump intensity factors are latent variables in the model, it is difficult to evaluate the likelihood function $\mathcal{L}$ in a direct manner. As such, we propose in this section a deterministic DNF scheme that uses a recursive prediction-update algorithm in the spirit of \citet{kitagawa1987non}. This methodology is based on a discretization of the state-space continuous latent variables, similar to the method proposed by \citet{fridman1998maximum}, \citet{bartolucci2001maximum}, \citet{clements2006mixture}, and \citet{langrock2012some}. It is important to note that such a methodology has not been applied to the likelihood evaluation of SVYJ, SVCJ and SVCJSI models, which is a key contribution of this paper. 



Let $\mathbf{x}_t \in \mathcal{X}$ be the latent state variables at time $t$ where $\mathbf{x}_t$ is either a scalar or a vector. Moreover, let us define by ${y}_t, t=1, 2, \dots, T$ the time-$t$ observations---i.e., returns in this study. Series of latent factors and observations are denoted by $\mathbf{x}_{0:t}  \equiv \{\mathbf{x}_s\}_{s=0}^t$ and $y_{1:t} \equiv \{y_s\}_{s=1}^t$. For a parameter set $\Theta$, the likelihood function of $\Theta$ is
\begin{align}
\mathcal{L}(\Theta) = &\, f\left( {y}_{1:T} \right) = f\left( {y}_{1}  \right) \, \prod_{t = 2}^T f\left( {y}_t \, \middle|\, {y}_{1:t-1}  \right) \label{eq:likelihood} =  \int_{\mathcal{X}^{T+1}} f \left( \mathbf{x}_{0:T}, {y}_{1:T} \right) \, d \mathbf{x}_{0:T} ,
\end{align}
which involves a multidimensional integration problem.

\subsection{General Methodology}
The one-step ahead \emph{prediction} of the latent states' distribution (conditional on the past returns) is given by
\begin{align*}
f\left( \mathbf{x}_t \, \middle|\, {y}_{1:t-1} \right) =&\, \int_{\mathcal{X}} q\left( \mathbf{x}_t \,\middle| \, \mathbf{x}_{t-1}, {y}_{1:t-1}\right) u \left( \mathbf{x}_{t-1} \,\middle|\, {y}_{1:t-1} \right) \, d\mathbf{x}_{t-1} ,
\end{align*}
where $q$ is the transition probability distribution of $\mathbf{x}_t$ given $\mathbf{x}_{t-1}$ and the past returns. Once a new return is available, the \emph{posterior} distribution of the state variables at time $t$, conditional on $\mathcal{G}_t$, may now be obtained as
\begin{equation}
u\left( \mathbf{x}_t \, \middle|\, {y}_{1:t} \right) = \frac{r\left( {y}_t \, \middle|\, \mathbf{x}_t, {y}_{1:t-1} \right) f\left( \mathbf{x}_t \, \middle|\, {y}_{1:t-1} \right) }{f\left( {y}_t \, \middle|\, {y}_{1:t-1} \right)} , \label{eq:update}
\end{equation}
where $r$ is the measurement probability density---or the conditional time-$t$ return density given the state space variables $\mathbf{x}_t$. This is also known as the \emph{update} step of the algorithm. The denominator of Equation~\eqref{eq:update} is the time-$t$ likelihood contribution of return $y_t$ conditional on the past returns, i.e.,
\begin{align*}
f\left( {y}_t \, \middle| \, {y}_{1:t-1}  \right) = &\, \int_{\mathcal{X}} r\left( {y}_t \, \middle|\, \mathbf{x}_t, {y}_{1:t-1} \right) f\left( \mathbf{x}_t \, \middle|\, {y}_{1:t-1} \right)\, d\mathbf{x}_t . \notag
\end{align*}


Given the model formulation of Section \ref{sec: models}, $\mathbf{x}_t$ depends upon $\mathbf{x}_{t-1}$. Hence, the time-$t$ likelihood contribution can be rewritten as
\begin{align}
 f\left({y}_{t} \, \middle|  \, {y}_{1:t-1}\right) = &\, \iint_{\mathcal{X}^2} r\left({y}_{t} \, \middle|  \, \mathbf{x}_t, \mathbf{x}_{t-1}, {y}_{1:t-1} \right) f\left(\mathbf{x}_{t}, \mathbf{x}_{t-1} \, \middle|  \, {y}_{1:t-1} \right) \, d \mathbf{x}_t \, d \mathbf{x}_{t-1} \notag \\
= &\, \iint_{\mathcal{X}^2} r\left({y}_{t} \, \middle|  \, \mathbf{x}_t, \mathbf{x}_{t-1}, {y}_{1:t-1} \right) q\left(\mathbf{x}_{t} \, \middle|  \, \mathbf{x}_{t-1} , {y}_{1:t-1} \right) u \left(\mathbf{x}_{t-1} \, \middle|  \, {y}_{1:t-1} \right) \, d \mathbf{x}_t \, d \mathbf{x}_{t-1},
\label{eq: like_GEN}
\end{align}
where $u \left(\mathbf{x}_{t-1} \, \middle|  \, y_{1:t-1} \right)$ is similar to Equation~\eqref{eq:update}. Then, the posterior density of Equation~\eqref{eq:update} can also be obtained recursively using the same idea:
\begin{align}
u \left(\mathbf{x}_{t} \, \middle|  \, y_{1:t} \right) = &\, \frac{\int_{\mathcal{X}} r\left({y}_{t} \, \middle|  \, \mathbf{x}_t, \mathbf{x}_{t-1}, {y}_{1:t-1} \right) f\left(\mathbf{x}_{t}, \mathbf{x}_{t-1} \, \middle|  \, {y}_{1:t-1} \right) \, d \mathbf{x}_{t-1}}{f\left(y_{t} \, \middle|  \, y_{1:t-1} \right)} \notag \\
= &\, \frac{\int_{\mathcal{X}}  r\left(y_{t} \, \middle|  \, \mathbf{x}_t, \mathbf{x}_{t-1}, y_{1:t-1} \right) q\left(\mathbf{x}_{t} \, \middle|  \, \mathbf{x}_{t-1} , y_{1:t-1} \right) u \left(\mathbf{x}_{t-1} \, \middle|  \, y_{1:t-1} \right) \,  d \mathbf{x}_{t-1}}{f\left(y_{t} \, \middle|  \, y_{1:t-1} \right)}. \label{eq:u_dnf}
\end{align}
Therefore, to obtain the likelihood function of Equation~\eqref{eq:likelihood}, one needs to find the posterior distribution of the latent states and the likelihood contribution at all times. Algorithm 1 summarizes the various steps of the general DNF.
\vspace{0.2cm}

\begin{algorithm}
\small
\caption{Discrete Nonlinear Filter}
\begin{algorithmic}[1]
  \STATE set $u\left( \mathbf{x}_0 \right) = p\left( \mathbf{x}_0 \right)$, where $p$ is the initial state density
  \FOR{each $t \in \{1, ..., T\}$}
  \STATE get the likelihood contribution $f\left( {y}_{t} \, \middle|  \, {y}_{1:t-1}\right)$ using numerical integration
  \STATE get the posterior distribution $u \left(\mathbf{x}_{t} \, \middle|  \, {y}_{1:t} \right)$ using numerical integration
  \ENDFOR
  \STATE compute the likelihood function $\mathcal{L}(\Theta)$ by taking the product of the likelihood contributions
\end{algorithmic}
\end{algorithm}
\vspace{-0.1cm}
\normalsize


\subsection{Likelihood Function of the SVCJSI Model}

We seek to derive the likelihood function of the SVCJSI model using the DNF method.\footnote{Being nested models, the SV, SVYJ and SVCJ models could also be estimated with the same method by reducing the problem's dimension.} Specifically, for the framework of Section~\ref{sec: models}, we have that $\mathbf{x}_t = \big[\, v_t\quad \lambda_t \quad  j_t^y \quad j_t^{\vphantom{y}v}  \, \big]$, where
\begin{equation}
j_t^y = \sum_{i=1}^{n_t} z_{t,i}^y \quad \text{and} \quad j_t^{\vphantom{y}v} = \sum_{i=1}^{n_t} z_{t,i}^v, \notag
\end{equation}
as knowing every individual jump sizes over a day is futile---only the aggregated jump is important in the discretization. Therefore, $\mathcal{X} = \mathbb{R}_+ \times \mathbb{R}_+ \times \mathbb{R} \times \mathbb{R}_+$ and the latent state vector is four-dimensional for the SVCJSI model.

Naively computing Equations~\eqref{eq: like_GEN} and \eqref{eq:u_dnf} might be cumbersome given the high-dimensionali\-ty of the state variables. Fortunately, some simplifications can be applied to reduce the dimension of the problem. First, by conditioning on the number of jumps at time $t$, we can simplify the transition density in the following way:
\begin{align*}
 q\left(\mathbf{x}_{t} \, \middle|  \, \mathbf{x}_{t-1} , {y}_{1:t-1}, n_t = n \right) = &\, q\left(v_t, \lambda_t,j_t^y, j_t^{\vphantom{y}v} \, \middle|  \, v_{t-1}, \lambda_{t-1}, j_{t-1}^y, j_{t-1}^{\vphantom{y}v}, {y}_{1:t-1}, n_t = n \right) \\
 = &\, q\left(v_t \, \middle|\, v_{t-1}, j_t^{\vphantom{y}v} \right) q \left( \lambda_t \, \middle|\, \lambda_{t-1} \vphantom{j_t^{\vphantom{y}v}} \right) q \left( \, j_{t}^{\vphantom{y}y} \, \middle|\, j_t^{\vphantom{y}v}, n_t = n \vphantom{j_t^{\vphantom{y}v}} \right)  q \left( \, j_{t}^{\vphantom{y}v} \, \middle|\,  n_t = n \vphantom{j_t^{\vphantom{y}v}} \right)
\end{align*}
as $v_t$ and $\lambda_t$ depend upon $v_{t-1}$ and $\lambda_{t-1}$, respectively. Yet, $j_t^{(y)}$ and $j_t^{(v)}$ do not depend on the previous value of the latent state; in fact, return and variance jump sizes are (iid) transient latent states. By conditioning again on the number of jumps at time $t$, the observation density is given by:
\begin{align*}
\, r\left(y_{t} \, \middle|  \, \mathbf{x}_t, \mathbf{x}_{t-1}, {y}_{1:t-1} , n_t = n \right) =  r\left(y_{t} \, \middle|  \, v_t, v_{t-1}, \lambda_t, \lambda_{t-1}, j_t^y, j_t^{\vphantom{y}v}, n_t = n \right).
\end{align*}
Then, following the rationale outlined at the beginning of Section~\ref{sec:dnf}, the time-$t$ likelihood contribution of Equation~\eqref{eq: like_GEN} can be rewritten as
\begin{align}
f\left(y_{t} \, \middle|  \, y_{1:t-1} \right) =& \, \iint_{\mathcal{X}^2} \! r\left({y}_{t} \, \middle|  \, \mathbf{x}_t, \mathbf{x}_{t-1}, {y}_{1:t-1} \right) q\left(\mathbf{x}_{t} \, \middle|  \, \mathbf{x}_{t-1} , {y}_{1:t-1} \right) u \left(\mathbf{x}_{t-1} \, \middle|  \, {y}_{1:t-1} \right) \, d \mathbf{x}_t \, d \mathbf{x}_{t-1} \notag  \\
=& \, \sum_{n=0}^{\infty} \iint_{\mathcal{X}^2} \! r\left({y}_{t} \, \middle|  \, \mathbf{x}_t, \mathbf{x}_{t-1}, {y}_{1:t-1}, n_t = n \right) q\left(\mathbf{x}_{t} \, \middle|  \, \mathbf{x}_{t-1} , {y}_{1:t-1}, n_t = n \right) \notag \\
&\, \hspace{1.60cm}  u \left(\mathbf{x}_{t-1} \, \middle|  \, {y}_{1:t-1} \right) \mathbb{P}\left( n_t = n \, \middle|\, \mathbf{x}_t, \mathbf{x}_{t-1} \right) \, d \mathbf{x}_t \, d \mathbf{x}_{t-1} \notag  \\
=&\, \sum_{n=0}^{\infty} \int\!\cdots\!\int_{\mathcal{X} \times \mathbb{R}_+ \times \mathbb{R}_+} \hspace{-0.25cm} \mathbb{P}\left( n_t = n  \, \middle|\, \lambda_{t-1}\right) r\left(y_{t} \, \middle|  \, v_t, v_{t-1}, \lambda_t, \lambda_{t-1}, j_t^y, j_t^{\vphantom{y}v}, n_t = n \right) \notag \\[-1ex]
& \hspace{3.1cm} q \left( \, j_{t}^{\vphantom{y}y} \, \middle|\, j_{t}^{\vphantom{y}v}, n_t = n \vphantom{j_t^{\vphantom{y}v}} \right) q\left(v_t \, \middle|\, v_{t-1}, j_t^{\vphantom{y}v} \right) q \left( \, j_{t}^{\vphantom{y}v} \, \middle|\,  n_t = n \vphantom{j_t^{\vphantom{y}v}} \right) q \left( \lambda_t \, \middle|\, \lambda_{t-1} \vphantom{j_t^{\vphantom{y}v}} \right) \notag \\
& \hspace{3.1cm}  u \left(v_{t-1}, \lambda_{t-1} \, \middle|  \, y_{1:t-1} \right)  \, d \mathbf{x}_t \, d v_{t-1} \, d \lambda_{t-1} \label{eq:likecontrib_inter}
\end{align}
because
\begin{equation}
\iint_{\mathbb{R}_+ \times \mathbb{R}} \hspace{-0.1cm} u \left(v_{t-1}, \lambda_{t-1},j_{t-1}^y, j_{t-1}^{\vphantom{y}v}  \, \middle|  \, y_{1:t-1} \right) \, dj_{t-1}^y \, dj_{t-1}^{\vphantom{y}v} =  u \left(v_{t-1}, \lambda_{t-1} \, \middle|  \, y_{1:t-1} \right). \notag
\end{equation}
Conditioning on the number of jumps also allows us to further simplify Equation~\eqref{eq:likecontrib_inter} so that
\begin{align}
 f\left(y_{t} \, \middle|  \, y_{1:t-1} \right) \notag =&\, \sum_{n=0}^{\infty} \int\!\cdots\!\int_{\mathbb{R}_+^5} \mathbb{P}\left( n_t \, \middle|\, \lambda_{t-1}\right) r\left(y_{t} \, \middle|  \, v_t, v_{t-1}, \lambda_t, \lambda_{t-1}, j_t^{\vphantom{y}v}, n_t = n \right)  \notag  \\[-1ex]
& \hspace{2.4cm} q\left(v_t \, \middle|\, v_{t-1}, j_t^{\vphantom{y}v} \right) q \left( \, j_{t}^{\vphantom{y}v} \, \middle|\,  n_t = n \vphantom{j_t^{\vphantom{y}v}} \right) q \left( \lambda_t \, \middle|\, \lambda_{t-1} \vphantom{j_t^{\vphantom{y}v}} \right)  \notag\\
& \hspace{2.4cm}  u \left(v_{t-1}, \lambda_{t-1} \, \middle|  \, y_{1:t-1} \right)  \, d v_{t} \, d \lambda_{t} \, d j_{t}^{\vphantom{y}v} \, d v_{t-1} \, d \lambda_{t-1} , \label{eq:final}
\end{align}
as
\begin{align*}
\int_{\mathbb{R}} r\left(y_{t} \, \middle|  \, v_t, v_{t-1}, \lambda_t, \lambda_{t-1}, j_t^y, j_t^{\vphantom{y}v}, n_t = n \right)  q \left( \, j_{t}^{\vphantom{y}y} \, \middle|\, n_t = n \vphantom{j_t^{\vphantom{y}v}} \right) \, d j_{t}^{\vphantom{y}y} =&\, r\left(y_{t} \, \middle|  \, v_t, v_{t-1}, \lambda_t, \lambda_{t-1}, j_t^{\vphantom{y}v}, n_t = n \right) , \notag
\end{align*}
which has a normal density. Specifically, we have that
\begin{equation}
\left. y_{t} \, \right| \, v_t, v_{t-1}, \lambda_t, \lambda_{t-1}, j_t^{\vphantom{y}v}, n_t = n \, \sim \! \mathcal{N} \left( \mu_t, \sigma_t^2 \right), \notag
\end{equation}
where
\begin{align*}
\mu_t =&\, \left( \mu - \frac{1}{2} v_{t-1} - \overline{\alpha} \lambda_{t-1} \right) h \!+\! \!\sqrt{v_{t-1}  h} \left( \rho_v \varepsilon_t^v + \rho_{\lambda}\varepsilon_t^{\lambda} \right) + \alpha n + \rho_z j_t^v , \\
\sigma_t^2 = &\, v_{t-1} \! \left( 1 \! -\! \rho_v^2 \!-\! \rho_{\lambda}^2 \right) h + n \delta^2,
\end{align*}
\begin{equation}
\varepsilon_t^v = \frac{v_t - v_{t-1} - \kappa(\theta-v_{t-1})h - j_t^{\vphantom{y}v}}{\sigma \sqrt{v_{t-1} h}} \quad \text{and} \quad
\varepsilon_t^{\lambda} = \frac{\lambda_t - \lambda_{t-1} - \chi(\theta-\lambda_{t-1})h}{\xi \sqrt{\lambda_{t-1} h}}. \notag
\end{equation}
We can obtain the integrand of Equation~\eqref{eq:final} in closed form because
\begin{align*}
\left.n_t\vphantom{j_{t}^{\vphantom{y}v}} \, \right| &\, \lambda_{t-1}  \,\,\hspace{0.45cm}\sim  \mathcal{P}(\lambda_{t-1} h), \\
\left.v_t\vphantom{j_{t}^{\vphantom{y}v}} \, \right|&\, v_{t-1}, j_t^{\vphantom{y}v}\,\, \sim \mathcal{N} \left( v_{t-1} + \kappa \left( \theta - v_{t-1} \right) h + j_t^{\vphantom{y}v}, \sigma^2 v_{t-1} h \right), \\
\left.j_{t}^{\vphantom{y}v} \right|&\,  n_t = n \,\,\hspace{0.08cm}\sim \Gamma(n, \nu), \\
\left.\lambda_t\vphantom{j_{t}^{\vphantom{y}v}} \, \right|&\, \lambda_{t-1}\,\, \hspace{0.48cm} \sim \mathcal{N} \left( \lambda_{t-1} + \chi \left( \omega - \lambda_{t-1} \right) h, \xi^2 \lambda_{t-1} h \right)
\end{align*}
and the corresponding densities are known in closed form as well. Note that $\mathcal{P}(m)$ is a Poisson distribution with mean $m$ and $\Gamma(n, \nu)$ is a gamma distribution with mean $n \nu$.


Finally, one can get the time-$t$ posterior \emph{density} of the latent states in a similar fashion by computing the following integral
\begin{align*}
&\, u\left( \mathbf{x}_{t} \, \middle|\, y_{1:t} \right) =  u\left( {v}_{t}, \lambda_t \, \middle|\, {y}_{1:t} \right) \\
= &\, \frac{1}{f\left( y_t \, \middle|\, y_{1:t-1} \right)} \Bigg( \sum_{n=0}^{\infty} \iiint_{\mathbb{R}_+^3} \mathbb{P}\left( n_t \, \middle|\, \lambda_{t-1}\right) r\left(y_{t} \, \middle|  \, v_t, v_{t-1}, \lambda_t, \lambda_{t-1}, j_t^{\vphantom{y}v}, n_t = n \right)  \\
&\, \hspace{3.30cm} q\left(v_t \, \middle|\, v_{t-1}, j_t^{\vphantom{y}v} \right) q \left( \, j_{t}^{\vphantom{y}v} \, \middle|\,  n_t = n \vphantom{j_t^{\vphantom{y}v}} \right) q \left( \lambda_t \, \middle|\, \lambda_{t-1} \vphantom{j_t^{\vphantom{y}v}} \right) u \left(v_{t-1}, \lambda_{t-1} \, \middle|  \, y_{1:t-1} \right)  \, d j_{t}^{\vphantom{y}v} \, d v_{t-1} \, d \lambda_{t-1} \Bigg).
\end{align*}

\subsection{Numerical Implementation}

To be able to numerically compute the integrals above, we need to define a discretization of the state space---so-called \emph{likely} sequences. The interval boundaries used to discretize the state space of the variance factor $v$, the jump intensity factor $\lambda$ and the variance jumps $j$, are chosen over the following ranges:
\begin{align*}
V \,\,=&\, \left[ \, v^{(1)} \quad \cdots \quad  v^{(N)} \,  \right], &  & & \Lambda \,\,=&\, \left[ \, \lambda^{(1)} \quad \cdots  \quad  \lambda^{(M)} \,  \right], & & &  {J} \,\,= &\, \left[ \, j^{(1)} \quad \cdots  \quad  j^{(K)} \,  \right],\\
v^{(1)} = &\, \, \mathcal{E}^V_{\infty} - \delta^V_N \sqrt{\mathcal{V}^V_{\infty}},& & & \lambda^{(1)} = &\,\, \mathcal{E}^{\Lambda}_{\infty} - \delta^{\Lambda}_M \sqrt{\mathcal{V}^{\Lambda}_{\infty}}, & & & j^{(1)} = &\,\, \mathcal{E}^{J}_{\infty} - \delta^{J}_K \sqrt{\mathcal{V}^{J}_{\infty}},\\
v^{(N)} = &\, \, \mathcal{E}^V_{\infty} + \delta^V_N \sqrt{\mathcal{V}^V_{\infty}},& & &\lambda^{(M)} = &\,\, \mathcal{E}^{\Lambda}_{\infty} +  \delta^{\Lambda}_M \sqrt{\mathcal{V}^{\Lambda}_{\infty}},& & &j^{(K)} = &\,\, \mathcal{E}^{J}_{\infty} +  \delta^{J}_K \sqrt{\mathcal{V}^{J}_{\infty}}, \\
\mathcal{E}^V_{\infty} = &\, \lim_{t\rightarrow \infty} \mathbb{E}^{\mathbb{P}}\left[ v_{t}\,\middle|\, \mathcal{F}_0 \right], & & & \mathcal{E}^{\Lambda}_{\infty} = &\, \lim_{t\rightarrow \infty} \mathbb{E}^{\mathbb{P}}\left[ \lambda_{t}\,\middle|\, \mathcal{F}_0 \right], & & & \mathcal{E}^{J}_{\infty} = &\, \lim_{t\rightarrow \infty} \mathbb{E}^{\mathbb{P}}\left[ j_{t}^{v} \,\middle|\, \mathcal{F}_0 \right],\\
\mathcal{V}^V_{\infty} = &\, \lim_{t\rightarrow \infty} \mathrm{Var}^{\mathbb{P}}\left[ v_{t}\,\middle|\, \mathcal{F}_0 \right], & & & \mathcal{V}^{\Lambda}_{\infty} = &\, \lim_{t\rightarrow \infty} \mathrm{Var}^{\mathbb{P}}\left[ \lambda_{t}\,\middle|\, \mathcal{F}_0 \right] , & & & \mathcal{V}^{J}_{\infty} = &\, \lim_{t\rightarrow \infty} \mathrm{Var}^{\mathbb{P}}\left[ j_t^v \,\middle|\, \mathcal{F}_0 \right],
\end{align*}
where $\mathcal{E}^V_{\infty}$ and $\mathcal{V}^V_{\infty}$ are the long-run expected value and variance of the stationary process of the variance factor, respectively. Moreover, $\mathcal{E}^{\Lambda}_{\infty}$, $\mathcal{V}^{\Lambda}_{\infty}$, $\mathcal{E}^{J}_{\infty}$ and $\mathcal{V}^{J}_{\infty}$ are defined analogously for the jump intensity factor and for the variance jumps, respectively. The intermediate points are determined by the following equations:
\begin{align*}
v^{(i)} = &\,  \left( \sqrt{v^{(1)}} + \left( \frac{i-1}{N-1}\right) \left( \sqrt{v^{(N)}} - \sqrt{v^{(1)}}\right) \right)^2 , \hspace{0.65cm} i = 2, ..., N-1,  \\
\lambda^{(l)} = &\, \left( \sqrt{\lambda^{(1)}} + \left( \frac{l-1}{M-1}\right) \left( \sqrt{\lambda^{(M)}} - \sqrt{\lambda^{(1)}}\right) \right)^2 , \quad l = 2, ..., M-1, \\
j^{(j)} = &\,  j^{(1)} + \left( \frac{j-1}{M-1}\right) \left( j^{(K)} - j^{(1)} \right)  , \hspace{2.2cm} j = 2, ..., K-1,
\end{align*}
making the variance (jump intensity) likely sequences uniformly distributed in the volatility (square root of the jump intensity) domain.


The functions $\delta^V_N$, $\delta^{\Lambda}_M$ and $\delta^{J}_K$ assure that the DNF converges and shall therefore respect two sets of conditions:
\begin{align*}
\lim_{N\rightarrow \infty} \delta^V_N =&\, \infty, & & & \lim_{M\rightarrow \infty} \delta^{\Lambda}_M =&\, \infty, & & & \lim_{K\rightarrow \infty} \delta^{J}_K =&\, \infty, \\
\lim_{N\rightarrow \infty} \frac{\delta^V_N}{N} =&\, \, 0, & & & \lim_{M\rightarrow \infty} \frac{\delta^{\Lambda}_M}{M} =&\, \, 0, & & & \lim_{K\rightarrow \infty} \frac{\delta^{J}_K}{K} =&\, \, 0.
\end{align*}
The first set of conditions makes sure that we cover the domain as the number of nodes increases. On the other hand, the second set of conditions ensures that the partition becomes finer and finer as the number of nodes increases. Obviously, a myriad of such functions exists. In this study, we use $\delta_N^V = 3 + \log(N)$, $\delta_M^{\Lambda} = 3 + \log(M)$ and $\delta_K^{J} = 3 + \log(K)$, although other functions could be used (as long as they satisfy the two sets of conditions).

Moreover, in the spirit of \citet{langrock2012some}, a special integration rule will be used in this study: broadly speaking, if $g_1$ and $g_2$ are two functions, then
\begin{equation}
\int_a^b g_1(x) g_2(x) \, dx \approx g_1(c) \int_a^b g_2(x)\, dx , \label{eq:quad}
\end{equation}
where $c$ is a representative point in $[a,b)$ (e.g., the midpoint). Given that we are integrating products of probability density functions (pdfs), this integral rule will allow us to rewrite the integral of Equation (\ref{eq:quad}) as $g_1(c) \left(G_2(b) - G_2(a) \right)$, where $G_2$ is the corresponding cumulative distribution function of the pdf $g_2$.\footnote{Other quadratures have been proposed in the literature. For instance, \citet{fridman1998maximum} use of Gauss-Legendre quadrature. \citet{bartolucci2001maximum} and \citet{clements2006estimating} use a simple numerical integral scheme based on the midpoint $c$, i.e.,
\begin{equation}
\int_a^b f_1(x) g_2(x) \, dx \approx (b-a) g_1(c) g_2(c). \notag
\end{equation}}



Then, to compute numerical integrals, it is often more convenient to define intervals. Therefore, let
\begin{align*}
V^{(i)} =&\, \left[ \frac{v^{(i-1)} + v^{(i)}}{2} , \frac{v^{(i)} + v^{(i+1)}}{2} \right), \hspace{0.6cm} i = 1, ..., N, \\
\Lambda^{(l)} =&\, \left[ \frac{\lambda^{(l-1)} + \lambda^{(l)}}{2} , \frac{\lambda^{(l)} + \lambda^{(l+1)}}{2} \right), \quad l = 1, ..., M, \\
J^{(j)} =&\, \left[ \frac{j^{(j-1)} + j^{(j)}}{2} , \frac{j^{(j)} + j^{(j+1)}}{2} \right), \hspace{0.4cm} j = 1, ..., K,
\end{align*}
be three different sets of intervals that cover $[0,\infty)$, where $v^{(0)} = - v^{(1)}$, $\lambda^{(0)} = - \lambda^{(1)}$, $j^{(0)} = - j^{(1)}$, $v^{(N+1)} = \infty$, $\lambda^{(M+1)} = \infty$, and $j^{(K+1)} = \infty$.

Finally, the time-$t$ likelihood contribution is approximated by
\begin{align}
 \widehat{f}\left(y_{t} \, \middle|  \, y_{1:t-1} \right) \notag \approx&\, \sum_{n=0}^{R} \sum_{i_t = 1}^{N} \sum_{i_{t-1} = 1}^{N} \sum_{j_t = 1}^{K} \sum_{l_t = 1}^{M} \sum_{l_{t-1} = 1}^{M}   r\left(y_{t} \, \middle|  \, v^{(i_t)}, v^{(i_{t-1})}, \lambda^{(l_t)}, \lambda^{(l_{t-1})}, j^{(j_t)}, n_t = n \right)  \notag  \\[-1.5ex]
& \hspace{4.25cm}\mathbb{P}\left( n_t = n \, \middle|\, \lambda^{(l_{t-1})}\right) q\left(v_t \in V^{(i_t)} \, \middle|\, v^{(i_{t-1})}, j^{(j_t)} \right) \notag  \\
& \hspace{4.25cm}  q \left( \, j_{t}^{\vphantom{y}v} \in J^{(j_t)} \, \middle|\,  n_t = n \vphantom{j_t^{\vphantom{y}v}} \right) q \left( \lambda_t \in \Lambda^{(l_{t})} \, \middle|\, \lambda^{(l_{t-1})} \vphantom{j_t^{\vphantom{y}v}} \right)  \notag\\
& \hspace{4.25cm}  u \left(v^{(i_{t-1})}, \lambda^{(l_{t-1})} \, \middle|  \, y_{1:t-1} \right) , \label{eq:approx_like}
\end{align}
where $R$ is the truncation level of the Poisson random variable. The time-$t$ posterior \emph{density} $u\left(v_t, \lambda_t \, \middle|\, y_{1:t} \right)$ is also estimated in a similar way:
\begin{align}
 \widehat{u}\left( v^{(i_t)}, \lambda^{(l_t)} \, \middle|  \, y_{1:t} \right) \notag \approx&\, \frac{1}{\widehat{f}\left(y_{t} \, \middle|  \, y_{1:t-1} \right)} \sum_{n=0}^{R} \sum_{i_{t-1} = 1}^{N} \sum_{j_t = 1}^{K} \sum_{l_{t-1} = 1}^{M} r\left(y_{t} \, \middle|  \, v^{(i_t)}, v^{(i_{t-1})}, \lambda^{(l_t)}, \lambda^{(l_{t-1})}, j^{(j_t)}, n_t = n \right)  \notag  \\[-1.5ex]
& \hspace{4.95cm}\mathbb{P}\left( n_t = n \, \middle|\, \lambda^{(l_{t-1})}\right) q\left(v_t \in V^{(i_t)} \, \middle|\, v^{(i_{t-1})}, j^{(j_t)} \right) \notag  \\
& \hspace{4.95cm}  q \left( \, j_{t}^{\vphantom{y}v} \in J^{(j_t)} \, \middle|\,  n_t = n \vphantom{j_t^{\vphantom{y}v}} \right) q \left( \lambda_t \in \Lambda^{(l_{t})} \, \middle|\, \lambda^{(l_{t-1})} \vphantom{j_t^{\vphantom{y}v}} \right)  \notag\\
& \hspace{4.95cm}  u \left(v^{(i_{t-1})}, \lambda^{(l_{t-1})} \, \middle|  \, y_{1:t-1} \right) . \notag
\end{align}

\section{Accuracy and Reliability}\label{sec:accuracy}

We now aim to assess the accuracy and reliability of the DNF. Specifically, we first verify that it converges to the correct likelihood value by comparing the log-likelihood obtained with the DNF to that of the SIR assuming that a large computing budget is available. This experiment will tell us how the DNF performs in a best-case scenario.


As explained in \citet{malik2011particle}, the SIR-based likelihood function is unbiased (see Section~\ref{secsm:pf} of the Supplementary Material for more details on the SIR method). Therefore, when a large computing budget is used, the SIR-based likelihood estimate should be very close to its true value with very little sampling error. Hence, it can be used as a reliable benchmark to assess the accuracy and reliability of the DNF.

For the SIR method, we use an increasing number of particles as a function of the model complexity. Specifically, we use 100,000 particles for the SV model, 250,000 particles for the SVYJ model and 1,000,000 particles for both the SVCJ and SVCJSI models. Then, we set the number of nodes in the variance (variance jump) grid to $N = 200$ ($K = 100$) for the SV, SVYJ and SVCJ models.\footnote{According to our results, a reliable rule of thumb is to set the number of nodes of the variance jump grid, $K$, to be two and three times fewer than the number of points of the variance.} As for the SVCJSI model, we set $N = M = 50$ and $K = 25$ due to memory size limits.\footnote{We used a graphics processing unit (GPU) to compute the likelihood function of the SVCJSI model with the DNF; see Section \ref{sec: motivation_GPU}.}



\subsection{Random Series}

In this first test, we compare the likelihood of 1,000 one-year daily series taken at random using parameters that are also randomly generated. The aim of this test is to determine the accuracy of the DNF for a wide range of possible paths and parameter sets.

Therefore, we generate parameter sets using the following bounds:
\vspace{-1cm}

\singlespacing
\begin{eqnarray*}
-0.20 \, \leq& \mu & \leq\, \hphantom{1}0.20 \\[-0.5ex]
0.00 \, \leq& \kappa & \leq\, 10.00 \\[-0.5ex]
0.00 \, \leq& \theta & \leq\, \hphantom{1}0.10 \\[-0.5ex]
0.10 \, \leq& \sigma & \leq\, \hphantom{1}1.00 \\[-0.5ex]
-0.95 \, \leq& \rho_v & \leq\, \hphantom{1}0.95 \\[-0.5ex]
0.00 \, \leq& \chi & \leq\, 50.00 \\[-0.5ex]
0.00 \, \leq& \omega & \leq\, 25.00 \\[-0.5ex]
0.10 \, \leq& \xi & \leq\, 10.00 \\[-0.2ex]
-\max\left[ 0.95, 1 - \rho_v^2\right] \, \leq& \rho_{\lambda} & \leq\,\,\, \,  \max\left[ 0.95, 1 - \rho_v^2\right] \\[-0.2ex]
-0.05 \, \leq& \alpha & \leq\, \hphantom{1}0.05 \\[-0.5ex]
0.00 \, \leq& \delta & \leq\, \hphantom{1}0.10 \\[-0.5ex]
-5.00 \, \leq& \rho_z & \leq\, \hphantom{1}5.00 \\[-0.5ex]
0.00 \, \leq& \nu & \leq\, \hphantom{1}0.03 .
\end{eqnarray*}
\doublespacing
\hspace{-0.95cm} These values are reasonable bounds for the parameters; they span multiple potential and realistic cases. Then, based on the simulated parameters, we generate one-year paths and we compute the log-likelihood using both methods. Specifically, we compare both estimates using the absolute percentage error (APE):
\begin{equation}
\text{APE} = 100 \% \, \left| \frac{\log\left(\mathcal{L}_{\text{DNF}}\left(\Theta\right) \right)- \log\left(\mathcal{L}_{\text{SIR}}\left(\Theta\right)\right) }{\log\left(\mathcal{L}_{\text{SIR}}\left(\Theta\right)\right) } \right|. \label{eq:ape}
\end{equation}

Table~\ref{tab:APE_Random} reports seven quantiles of the APE distribution for the four models considered. Broadly speaking, the medians hover between 0.01\% and 0.05\% for the four models which is very small. The right tail of the APE distributions is also very thin with 99.5$^{\text{th}}$ quantiles that are between 0.40\% and 0.85\%.

\begin{table}[ht!]
{\footnotesize
\topcaption{\textbf{Distribution of the Log-Likelihood Absolute Percentage Error: Random Series.}}
{\centering
\begin{tabularx}{\linewidth}{lCCCC}
\toprule
$\alpha$ & \textbf{SV} & \textbf{SVYJ} & \textbf{SVCJ} & \textbf{SVCJSI} \\
\cmidrule(lr){2-2} \cmidrule(lr){3-3} \cmidrule(lr){4-4} \cmidrule(lr){5-5}
{0.250} & 0.0039 & 0.0048 & 0.0133 & 0.0159 \\
{0.500} & 0.0088 & 0.0123 & 0.0512 & 0.0367 \\
{0.750} & 0.0168 & 0.0263 & 0.1619 & 0.0843 \\
{0.900} & 0.0318 & 0.0530 & 0.3608 & 0.1931 \\
{0.950} & 0.0545 & 0.0781 & 0.5016 & 0.3773 \\
{0.990} & 0.1986 & 0.2108 & 0.6838 & 0.7536 \\
{0.995} & 0.4318 & 0.5068 & 0.7235 & 0.8247 \\
\bottomrule
\end{tabularx}}
\label{tab:APE_Random}
}
\footnotesize

This table reports the quantiles of the absolute percentage error distribution for the four models considered in this study. These distributions are obtained by generating 1,000 one-year series using randomly selected parameters and by computing the APE for each path as shown in Equation~\eqref{eq:ape}. SV stands for stochastic volatility, SVYJ for stochastic volatility with return jumps, SVCJ for stochastic volatility with correlated and simultaneous jumps in return and variance, and SVCJSI for SVCJ with stochastic jump intensity.
\end{table}
\normalsize

\begin{figure}[ht!]
\centering
\includegraphics[width=0.9\linewidth]{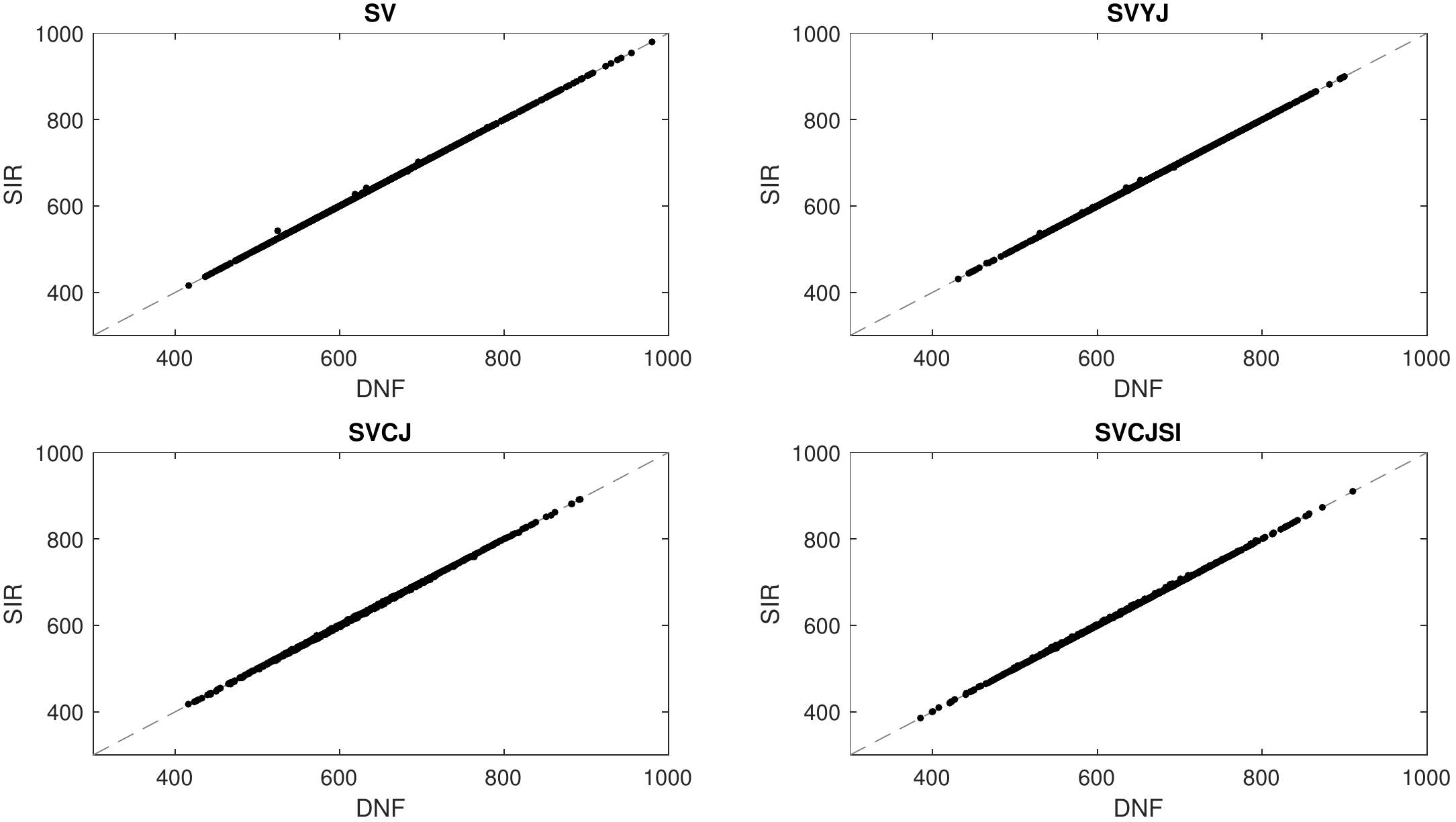}
\caption{{\bf SIR- Against DNF-Based Log-Likelihood Estimates: Random Series.} \newline \hspace{10cm}
\footnotesize
This figure reports the SIR- and DNF-based log-likelihood estimates for large computing budget and for the four models considered in this study. These scatter plots are obtained by generating 1,000 one-year series using randomly selected parameters and by computing the log-likelihood for each path by using both the DNF and the SIR methods.}
\label{fig:Scatter_Random}
\end{figure}

Figure~\ref{fig:Scatter_Random} supplements Table~\ref{tab:APE_Random} by reporting scatter plots of SIR- and DNF-based log-likeli\-hoods. Most of these are aligned on the diagonal meaning that both methods are yielding very much alike estimates when using a large computing budget. Based upon this first test, we can comfortably conclude that the DNF is very accurate, at least within the range of parameters provided above and for a large computing budget.

\subsection{S\&P 500 Index\label{subsubsec:SP500}}

Instead of using randomly generated series that might not resemble those used in empirical finance, we now focus on the S\&P 500 returns as these would be representative of typical asset returns. Specifically, we use daily S\&P 500 index returns (excluding dividends), from January 1990 to September 2018, obtained from the Bloomberg terminal. For this specific time series, we then randomly generate a set of parameters and compute the resulting likelihood with both the SIR and DNF methods.


\begin{table}[!ht]
\footnotesize
\topcaption{\textbf{Distribution of the Log-Likelihood Absolute Percentage Error: S\&P 500 Index.}}
{\centering
\begin{tabularx}{\linewidth}{lCCCC}
\toprule
$\alpha$ & \textbf{SV} & \textbf{SVYJ} & \textbf{SVCJ} & \textbf{SVCJSI} \\
\cmidrule(lr){2-2} \cmidrule(lr){3-3} \cmidrule(lr){4-4} \cmidrule(lr){5-5}
0.250 & 0.0118 & 0.0161 & 0.0122 & 0.0403 \\
0.500 & 0.0245 & 0.0358 & 0.0311 & 0.1020 \\
0.750 & 0.0513 & 0.0712 & 0.0659 & 0.2305 \\
0.900 & 0.1049 & 0.1155 & 0.1152 & 0.4243 \\
0.950 & 0.1793 & 0.2025 & 0.1477 & 0.5971 \\
0.990 & 0.3123 & 0.5546 & 0.2991 & 0.7696 \\
0.995 & 0.3503 & 0.6336 & 0.4236 & 0.7812 \\
\bottomrule
\end{tabularx}}
\label{tab:APE_SP500}
\footnotesize

This table reports the quantiles of the absolute percentage error distribution for the four models considered in this study. These distributions are obtained by generating 1,000 sets of parameters and by computing the APE for each set as shown in Equation~\eqref{eq:ape} with the S\&P 500 Index time series.
\end{table}
\normalsize

Table~\ref{tab:APE_SP500} mimics the results of Table~\ref{tab:APE_Random}, only this time we consider the S\&P 500 returns over the last three decades. The results are similar to those obtained with random series: in fact, the median APE is rather small, ranging between 0.02\% and 0.10\%. Again, even the worst cases are still very decent,  with the 99.5$^{\text{th}}$ quantile being below 1\% for the four models. Therefore, for a typical financial time series and a large computing budget, the DNF is very accurate.

\section{Precision and Computing Times}
\label{sec:speed}
As shown in the previous section, the DNF is capable of providing very accurate log-likelihood estimates when a large computing budget is available. Nonetheless, it is of great interest to also assess its performance when a smaller budget is considered because most inference methods---both frequentist and Bayesian---require a considerable number of likelihood function evaluations. The overarching goal of this section is to analyze the precision of the DNF compared with computing times required to attain such precision.

We first discuss the use of central processing unit (CPU) and graphics processing unit (GPU) for the problem at hand. We then compare the DNF and SIR in terms of computing times and precision. Finally, we analyze the speed and accuracy trade-off of the DNF by determining the grid size or computing time necessary to achieve a given level of precision (and vice versa).


\subsection{Motivation for CPU and GPU Computing}\label{sec: motivation_GPU}
The computing times involved in this study are greatly influenced by the number of state variables considered, which has an impact on the dimension of the sum of Equation~\eqref{eq:approx_like}. Indeed, Equation~\eqref{eq:approx_like} involves a sum over two, three, four or six dimensions for the SV, SVYJ, SVCJ and SVCJSI models, respectively. Obviously, such calculations can become cumbersome when $N$, $M$, $K$, $R$ and $T$ are large.

It is easy to see why a GPU can be very useful. For instance, if $N = 50$, $M=50$, $K = 25$ and $R = 2$, then the size of the six-dimensional matrix containing all the elements that need to be summed up is
\begin{align*}
50 \times 50 \times 50 \times 50 \times 25 \times 2 \times 8 \text{ bytes } &\, = 2{,}500{,}000{,}000 \text{ bytes }  = 2.5 \text{ gigabytes.}
\end{align*}
Moreover, the number of operations needed to obtain one likelihood contribution is rather large and it increases exponentially, e.g., $5 \times 50 \times 50 \times 50 \times 50 \times 25 \times 2 = 1{,}562{,}500{,}000$ multiplications, and $50 \times 50 \times 50 \times 50 \times 25 \times 2 = 312{,}500{,}000$ additions for the example above.

\begin{figure}
\centering
\includegraphics[width=0.9\linewidth]{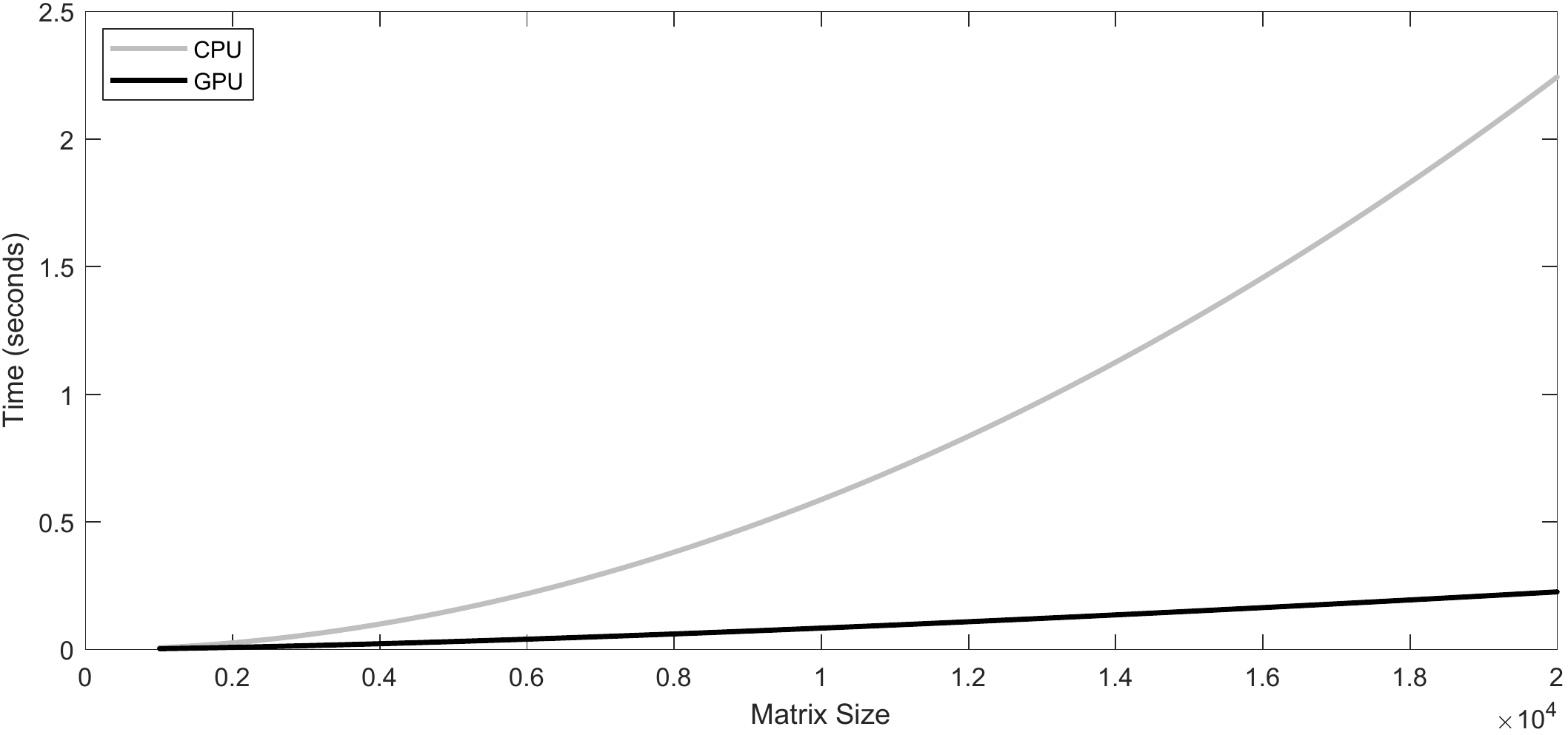}
\caption{{\bf Computing Time Against Matrix Size: GPU versus CPU.} \newline
\footnotesize
This figure reports the time needed to compute the sum of all the entries of an $n \times n$ matrix, where $n$ is the matrix size, as a function of a matrix size. Central processing unit is denoted by CPU and graphics processing unit by GPU.}
\label{fig:CPUvsGPU}
\end{figure}

CUDA-enabled GPUs are extremely efficient when it comes to calculating sums. In fact, for very large matrices, the time needed to compute sums is manifold lower when using a GPU instead of a CPU. Figure~\ref{fig:CPUvsGPU} shows the computing time for summing all the entries of an $n \times n$ matrix while using a GPU (black) or a CPU (grey). For instance, the GPU-based calculation is eight times faster for a matrix of size 20{,}000 $\times$ 20{,}000---which is a significant decrease.

But for GPU computing to cut overall computing times, it is important to take into account the additional time required to move data between the CPU and GPU, known as the overhead. For array sizes involved with deterministic jump arrival intensity models, our experiments showed that a conventional CPU is still faster overall.

Therefore and unless stated otherwise, the SIR and DNF methods have been coded in Matlab 2018b on a single thread with a typical CPU. However, the DNF method applied to the SVCJSI model has been implemented with an NVIDIA QUADRO P6000 GPU, also using Matlab 2018b.\footnote{The computing times are obtained by running Matlab programs on a desktop computer with two 2.1 GHz Intel Xeon E5-2620 v4 and 48 GB RAM. The NVIDIA QUADRO P6000 we use has 3840 CUDA parallel-processing cores and a GPU memory of 24 GB.}

\subsection{Deterministic Jump Arrival Intensity: SV, SVYJ and SVCJ Models}\label{sec: SV-SVYJ-SVCJ_models}

Since models with deterministic and stochastic jump arrival intensity have been implemented differently, we analyze computing times and precision for both families of models separately.

\subsubsection{Computing Times: Comparison Between DNF and SIR}

The goal of this first test is to analyze the accuracy of SIR and DNF for different fixed and finite computing budgets. For the SIR method, the computing budget is determined by the number of particles whereas for the DNF method, the grid size parameters $N$ and $K$ determine the computing budget. For all computations, we use the same daily S\&P 500 return series over 5 years (about 1250 observations) evaluated with the parameters that produced the highest likelihood in Section~\ref{subsubsec:SP500} as a proxy for maximum likelihood parameters.


Then, for a given computing budget, we compute the likelihood value and record the computing time. Moreover, for the SIR, we repeat the exercise 50 times to approximate the distribution of the log-likelihood values. This entire exercise is then rerun for several computing budgets. 


\begin{figure}[!ht]
\centering
\includegraphics[width=0.9\linewidth]{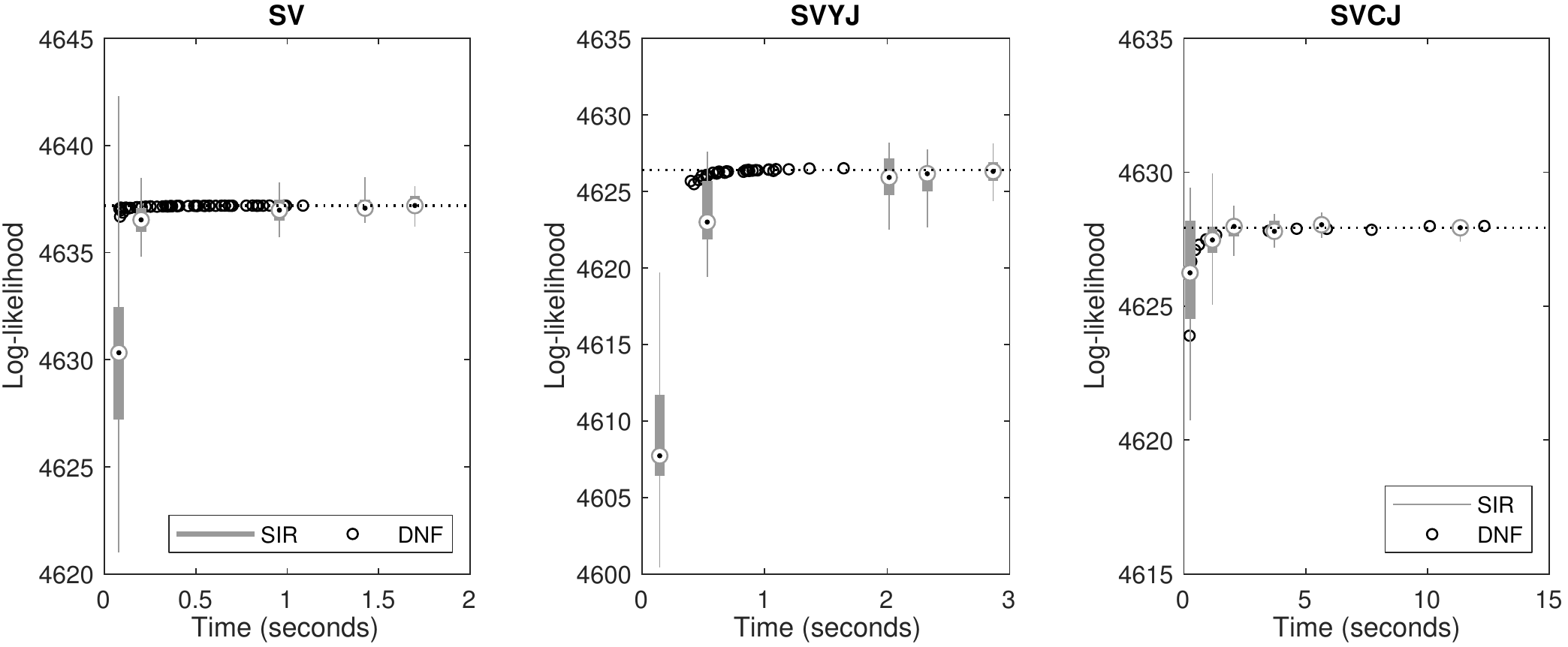}
\caption{{\bf DNF-Based Log-Likelihood estimates and the Distribution of the SIR-Based Log-Likelihood Estimates Against Time.} \newline \hspace{10cm}
\footnotesize
This figure reports the DNF-based log-likelihood estimates (black circles) and the distribution of the SIR-based log-likelihood estimates (grey box plots), while using small computing budgets. We use the same daily S\&P 500 return series over a five year period. For the SIR, we repeat the exercise 50 times to get a distribution of the log-likelihood value. The size of the budget is proxied by the computing time---the number of seconds it took to run one log-likelihood evaluation. We consider boxplots with whiskers from minimum to maximum and boxes from the lower quartile to the upper one. A circle with a dot represents the median of the SIR distribution.}
\label{fig:SmallerBudget1}
\end{figure}

Figure~\ref{fig:SmallerBudget1} reports this comparison. We show boxplots (in grey) of the distribution of the SIR-based log-likelihood values and we compare them to values obtained with the DNF method (black circles). The \emph{true} value (dashed line) is given by the average obtained with SIR using a very large computing budget (this number is virtually the same as that obtained with the DNF method for a large budget as well).

For the SV model, the DNF method yields values that are very close to the true log-likelihood value---and this, after only 100 milliseconds (ms). For such small computing budget, the SIR does not produce adequate results: it is biased and shows important variations. For a budget of one second per likelihood valuation, the SIR still gives very noisy log-likelihood estimates even though this value is asymptotically consistent: the whiskers are quite long, but the median is aligned with the true value.

The DNF provides similar results for the SVYJ model: after 600 ms, the deterministic method produces very steady and stable estimates of the log-likelihood function---consistent with the true value obtained with a large computing budget. The SIR-based log-likelihood, on the other hand, is still volatile even with a budget of two seconds.

The results for the SVCJ model tell a similar story: after a few seconds, the DNF is precise and stable and quickly converges to the true value whereas the SIR-based log-likelihood still shows significant sampling variations as the whiskers are still clearly visible to the naked eye.

\subsubsection{Speed and Accuracy Trade-Off}
We know the DNF method performs very well compared to the SIR providing accurate results much quicker than the SIR. In this second test, we would like to determine the precision gained (or lost) if we have a limited computing budget, as defined by either the grid size or computing time.

In this experiment, we generate 100 sets of parameters and we evaluate the likelihood function for varying number of nodes. We use the number of variance nodes $N$ ranging from 25 to 200 with steps of 5. The number of variance jump nodes is consistent with the rule of thumb explained above, i.e., $N/2.5$. The number of nodes associated with the jumps, $R$, is set to 2. We still use daily S\&P 500 returns over 5 years for about 1250 observations.

The likelihood values obtained for these varying number of nodes is then compared to that obtained with the SIR while using a large computing budget (see Section~\ref{sec:accuracy}). We still use the APE as a basis of comparison and compute mean APE (MAPE) to assess the average difference between the log-likelihood values:
\begin{equation}
\text{MAPE} = 100 \% \left( \frac{1}{100} \, \sum_{i= 1}^{100} \left| \frac{\log\left(\mathcal{L}_{\text{DNF}}\left(\Theta_i\right) \right)- \log\left(\mathcal{L}_{\text{SIR}}\left(\Theta_i\right)\right) }{\log\left(\mathcal{L}_{\text{SIR}}\left(\Theta_i\right)\right) } \right| \right). \notag
\end{equation}

\begin{figure}[!ht]
\centering
\includegraphics[width=0.9\linewidth]{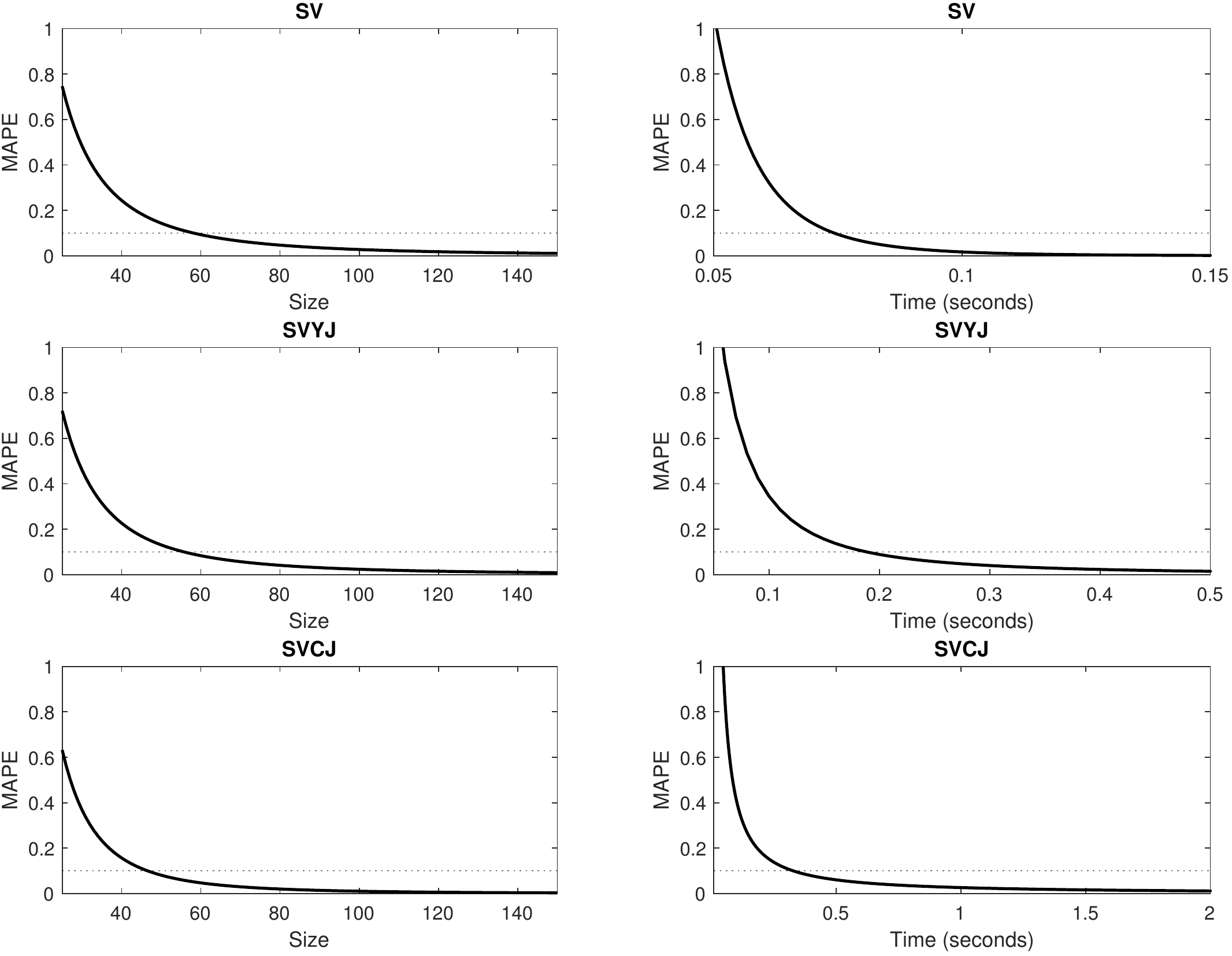}
\caption{{\bf Mean Absolute Percentage Error Against the Number of Variance Nodes and the Computing Time.} \newline \hspace{10cm}
\footnotesize
This figure reports the smoothed MAPE per model against the size---proxied as the number of variance nodes---on the leftmost panels and against computing time (in seconds) on the rightmost panels. We use daily returns from the S\&P 500 over 5 years, generate 100 sets of parameters and we evaluate the likelihood function for varying number of nodes. We use number of variance nodes $N$ ranging from 25 to 200 with steps of 5. The number of variance jump nodes is consistent with the rule of thumb explained above, i.e., $N/2.5$. The number of nodes associated with the jumps, $R$, is set to 2 in all of our tests.
}
\label{fig:TradeOff1}
\end{figure}

The leftmost panels of Figure~\ref{fig:TradeOff1} report the smoothed version of the MAPE against size---here, proxied by the number of nodes in the variance grid $N$.\footnote{We use Matlab's \texttt{fit} function with a power model, i.e., $y = a x^b$, to fit the MAPE curve. The results are robust to other specifications.} For the four models considered, a size between 50 and 60 would lead to a MAPE below 0.1\%. This result is in fact consistent with the size used in \citet{watanabe1999non}, \citet{clements2004discretised}, \citet{clements2006estimating} and \citet{langrock2012some}.

The rightmost panels of Figure~\ref{fig:TradeOff1} detail the relationship between the MAPE and the computing time per likelihood valuation (in seconds). Again, as explained above, the computing time tends to increase with the complexity of the model. It takes approximately 75 ms, 200 ms and 300 ms to attain an MAPE of about 0.1\% with the SV, SVYJ and SVCJ models, respectively. Therefore, with a conventional CPU, one is able to compute the likelihood function in much less than a second and reach an MAPE of 0.1\%.



\subsection{Stochastic Jump Arrival Intensity: SVCJSI Model}
We now analyze the performance of the DNF for the SVCJSI model. As opposed to the family of deterministic jump arrival intensity models, the SVCJSI model is comprised of four state variables and to accelerate computations, we implemented the DNF using a high-end GPU.

\begin{figure}[!ht]
\centering
\includegraphics[width=0.48\linewidth]{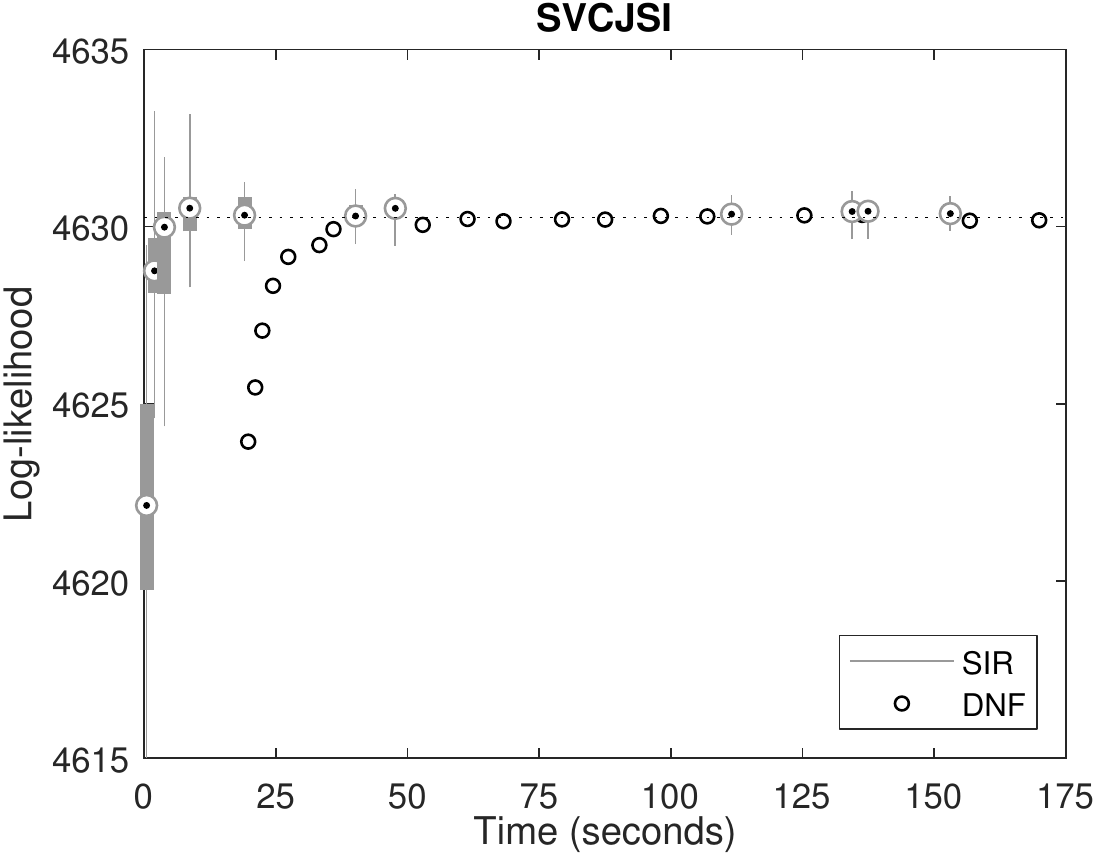}
\caption{{\bf DNF-Based Log-Likelihood estimates and the Distribution of the SIR-Based Log-Likelihood Estimates Against Time, continued.} \newline \hspace{10cm}
\footnotesize
See the description of Figure~\ref{fig:SmallerBudget1} for more details.}
\label{fig:SmallerBudget2}
\end{figure}

As in Section \ref{sec: SV-SVYJ-SVCJ_models}, we first seek to compare the precision and computing times of the SIR and DNF methods. Figure~\ref{fig:SmallerBudget2} is the same as Figure~\ref{fig:SmallerBudget1}, but for the SVCJSI model.

For a very small computing budget (a couple of seconds), the SIR generates a very wide range of likelihood values. As we add particles and thus the required computing time, the width of whiskers decreases as expected. However, even if the computing budget is large (2 or 3 minutes per likelihood valuation), there is still a significant uncertainty that remains around the likelihood value obtained. But if we are equipped with a powerful GPU, one is capable of achieving a very precise likelihood value with the DNF within a minute.


Figure~\ref{fig:SmallerBudget2} also illustrates the difficulty of precisely evaluating the likelihood function for the SVCJSI model (with the technology available to most users in 2019). If the computing budget available is small (e.g., a couple of seconds), then one should turn to the SIR implemented on a CPU but the resulting likelihood value will be biased and random. If a GPU is available, the computing budget must be large enough to overcome the extra overhead: at least 30 to 60 seconds are needed to achieve a level of accuracy that is impossible to attain with the SIR with a finite computing budget.



\begin{figure}[!ht]
\centering
\includegraphics[width=0.9\linewidth]{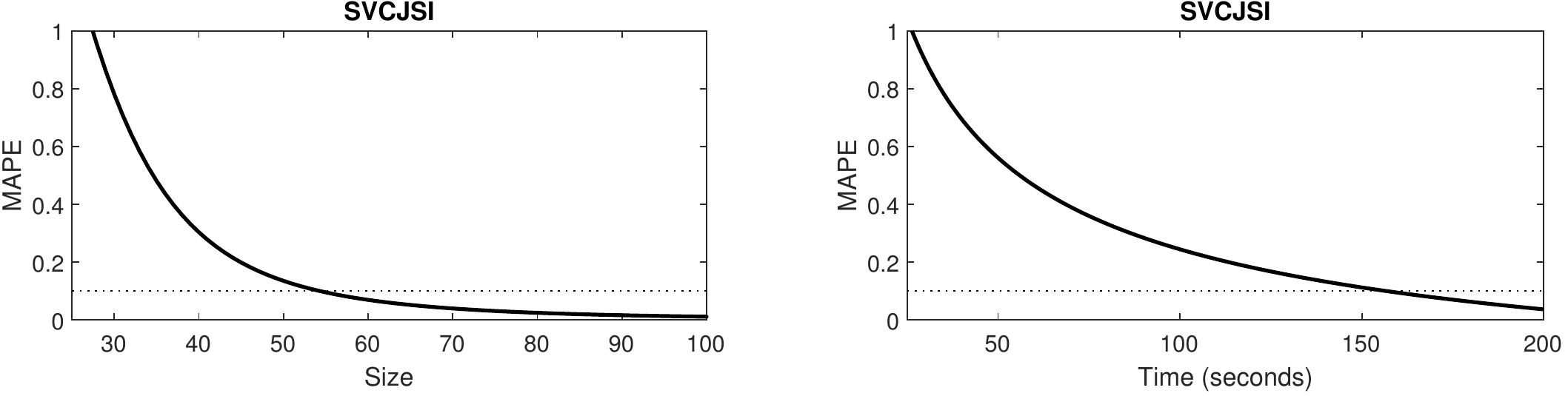}
\caption{{\bf Mean Absolute Percentage Error Against the Number of Variance Nodes and the Computing Time, continued.} \newline \hspace{10cm}
\footnotesize
This figure reports the smoothed MAPE per model against the size---proxied as the number of variance nodes---on the leftmost panels and against computing time (in seconds) on the rightmost panels. We use daily returns from the S\&P 500 over 5 years and generate 100 sets of parameters and we evaluate the likelihood function for varying number of nodes. For the SVCJSI model, the number of variance and jump intensity nodes ranges from 25 to 50 with steps of 1, with $N=M$. The number of variance jumps is set to $\lceil N/2.5 \rceil$ and the number of jump nodes is set to 2 still.
}
\label{fig:TradeOff2}
\end{figure}

We also analyze the MAPE as a function of the grid size or computing time. The number of variance and jump intensity nodes ranges from 25 to 50 with steps of 1, with $N=M$.\footnote{A number of nodes of 50 is the physical size limit the GPU handles at the time of writing this article.} Moreover, the number of variance jumps is set to $\lceil N/2.5 \rceil$ and the number of jump nodes is set to 2 still.

The results are illustrated in Figure~\ref{fig:TradeOff2} for the SVCJSI model. Again, we observe that a grid size of about 50-60 achieves an MAPE of 0.1\% (as with the SV, SVYJ and SVCJ models) but the required computing time to achieve such MAPE is about 160 seconds with GPU computing.

\section{Empirical Applications\label{sec:empirical}}

In the previous applications, we have shown the precision of the DNF in evaluating the likelihood function. For frequentist-based inference such as the maximum likelihood, the likelihood function must be smooth enough for an optimizer to find a global maximum in the parameter space. As such, the DNF has a key advantage over the SIR as its construction leads to a smooth likelihood function.



We thus present in this last section an important application of the DNF methodology: we compute the maximum likelihood estimates (MLE) of each of the four models studied in this paper. This section is certainly one of the few attempts in the literature to compute the MLE of such complex financial models.

\begin{figure}[ht!]
\centering
\includegraphics[width=0.9\linewidth]{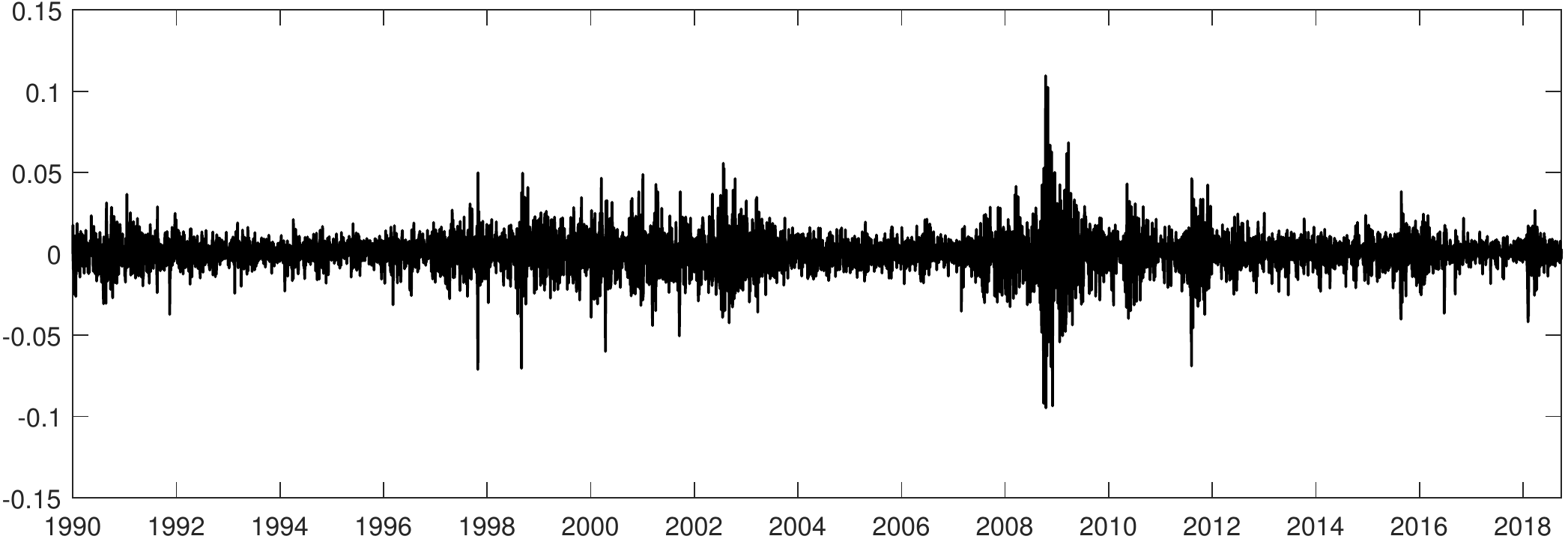}
\caption{{\bf S\&P 500 Index Returns (1990--2018).} \newline \hspace{10cm}
\footnotesize
This figure reports daily S\&P 500 index returns (excluding dividends), from January 1990 to September 2018, obtained from the Bloomberg terminal.}
\label{fig:SP500_Data}
\end{figure}

We use the S\&P 500 return series (excluding dividends) from 1990 to 2018 (see Figure~\ref{fig:SP500_Data}). This time frame contains a few recessions and eras of turmoil: the early 1990s recession (subsequent to the 1990 oil price shock), the early 2000s recession (partly caused by the collapse of the speculative dot-com bubble and the 9/11 attacks) and the Great Recession (a by-product of the collapse of the US housing bubble, followed by a global financial crisis). This sample should therefore contain a time-varying volatility, return jumps, variance jumps, and potentially, a changing jump intensity. The grid sizes used for the computing of the likelihood function are such that $N = M = 50$, $K = 20$ and $R = 2$, which are consistent with our results.

Table~\ref{tab:SP500_Parameters} reports the MLE for each model along with their standard error in brackets.\footnote{Robust standard errors (in brackets) are calculated from the outer product of the gradient at the optimum parameter value. For more details on this method, see Appendix B.5 of \citet{remillard2013statistical}.} These estimates are comparable to those found in other studies and using different statistical methodologies \citep[e.g.,][]{andersen2002empirical,eraker2003impact,christoffersen2010volatility,hurn2015estimating}.


\begin{table}[ht!]
\footnotesize
\topcaption{\textbf{Estimated Parameters for the S\&P 500 Index.}}
{\centering
\begin{tabularx}{\linewidth}{lCCCC}
\toprule
      & \textbf{SV} & \textbf{SVYJ} & \textbf{SVCJ} & \textbf{SVCJSI} \\
      \cmidrule(lr){2-2} \cmidrule(lr){3-3} \cmidrule(lr){4-4} \cmidrule(lr){5-5}
$\mu$    & \hspace{0.21cm}0.041 & \hspace{0.21cm}0.035 & \hspace{0.21cm}0.038 & \hspace{0.21cm}0.035 \\
      & \hspace{0.21cm}(0.017) & \hspace{0.21cm}(0.016) & \hspace{0.21cm}(0.020) & \hspace{0.21cm}(0.024) \\
$\kappa$ & \hspace{0.21cm}5.923 & \hspace{0.21cm}6.357 & \hspace{0.21cm}3.689 & \hspace{0.21cm}4.316 \\
      & \hspace{0.21cm}(0.405) & \hspace{0.21cm}(0.343) & \hspace{0.21cm}(0.311) & \hspace{0.21cm}(0.201) \\
$\theta$ & \hspace{0.21cm}0.031 & \hspace{0.21cm}0.027 & \hspace{0.21cm}0.032 & \hspace{0.21cm}0.034 \\
      & \hspace{0.21cm}(0.002) & \hspace{0.21cm}(0.001) & \hspace{0.21cm}(0.002) & \hspace{0.21cm}(0.002) \\
$\sigma$ & \hspace{0.21cm}0.514 & \hspace{0.21cm}0.488 & \hspace{0.21cm}0.446 & \hspace{0.21cm}0.452 \\
      & \hspace{0.21cm}(0.017) & \hspace{0.21cm}(0.010) & \hspace{0.21cm}(0.018) & \hspace{0.21cm}(0.001) \\
$\rho_v$ & $-$0.692 & $-$0.708 & $-$0.745 & $-$0.666 \\
      & \hspace{0.21cm}(0.024) & \hspace{0.21cm}(0.024) & \hspace{0.21cm}(0.024) & \hspace{0.21cm}(0.031) \\
$\chi$   & --    & --    & --    & \hspace{0.21cm}2.706 \\
      & --    & --    & --    & \hspace{0.21cm}(2.618) \\
$\omega$ & --    & \hspace{0.21cm}2.487 & \hspace{0.21cm}5.125 & \hspace{0.21cm}3.232 \\
      & --    & \hspace{0.21cm}(1.487) & \hspace{0.21cm}(3.021) & \hspace{0.21cm}(3.111) \\
$\xi$    & --    & --    & --    & \hspace{0.21cm}6.947 \\
      & --    & --    & --    & \hspace{0.21cm}(5.778) \\
$\rho_{\lambda}$ & --    & --    & --    & $-$0.411 \\
      & --    & --    & --    & \hspace{0.21cm}(0.102) \\
$\alpha$ & --    & $-$0.014 & $-$0.007 &$-$0.014 \\
      & --    & \hspace{0.21cm}(0.007) & \hspace{0.21cm}(0.005) & \hspace{0.21cm}(0.016) \\
$\delta$ & --    & \hspace{0.21cm}0.008 & \hspace{0.21cm}0.003 & \hspace{0.21cm}0.005 \\
      & --    & \hspace{0.21cm}(0.003) & \hspace{0.21cm}(0.005) & \hspace{0.21cm}(0.013) \\
$\nu$    & --    & --    & \hspace{0.21cm}0.004 & \hspace{0.21cm}0.011 \\
      & --    & --    & \hspace{0.21cm}(0.001) & \hspace{0.21cm}(0.006) \\
$\rho_z$ & --    & --    & $-$1.809 & $-$1.381 \\
      & --    & --    & \hspace{0.21cm}(0.671) & \hspace{0.21cm}(0.706) \\
\bottomrule
\end{tabularx}}
\label{tab:SP500_Parameters}
\footnotesize

This table reports the maximum likelihood estimates for the four models considered in this study. We use the S\&P 500 return series from 1990 to 2018. Robust standard errors (in brackets) are calculated from the outer product of the gradient at the optimum parameter value.
\end{table}
\normalsize

\begin{sidewaysfigure}
\centering
\includegraphics[width=0.9\linewidth]{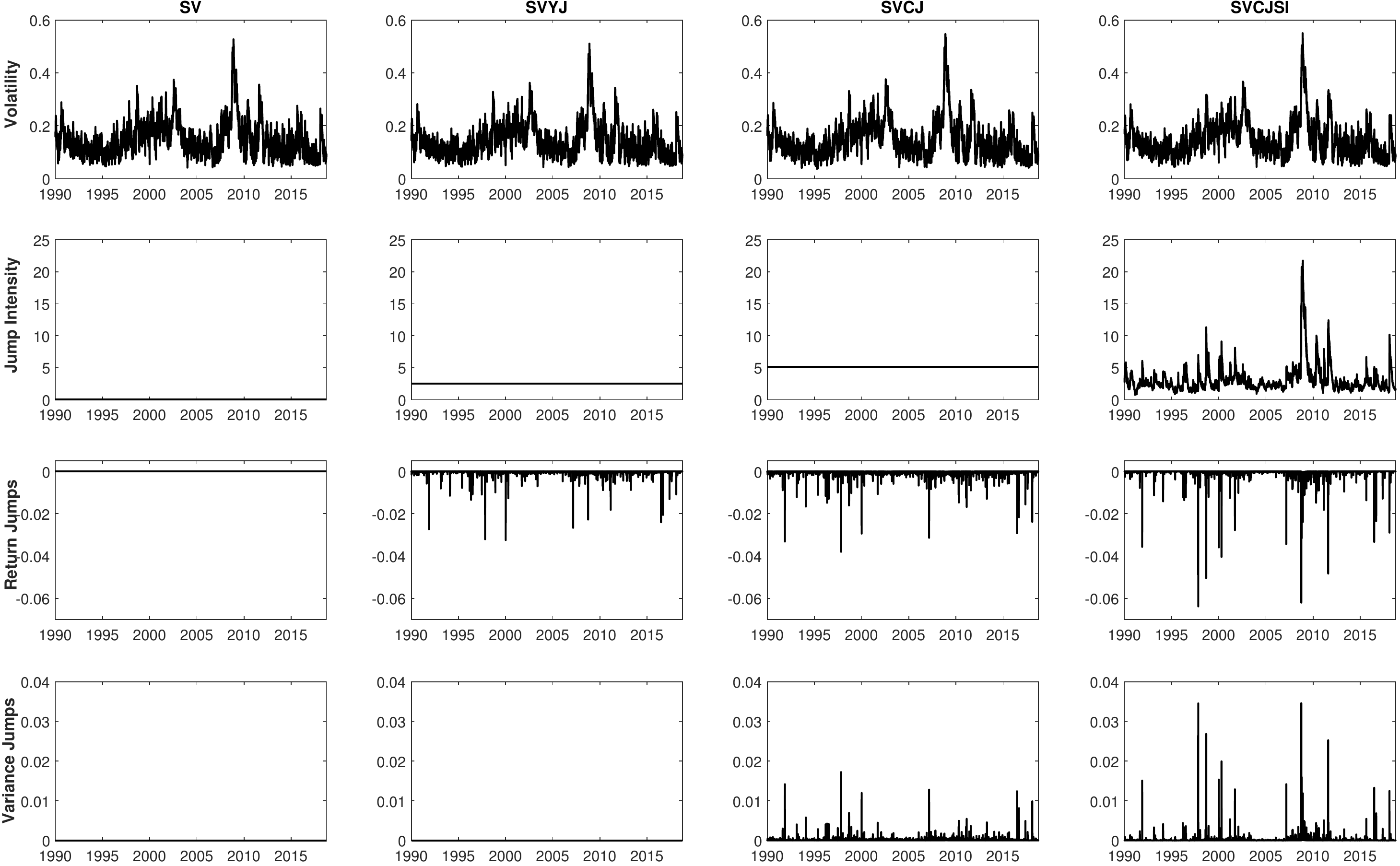}
\caption{{\bf Filtered Volatility, Jump Intensity, Return Jumps and Variance Jumps for the S\&P 500 Index.} \newline \hspace{10cm}
\footnotesize
This figure reports the filtered volatility (i.e., $\sqrt{V_t/252}$), the jump intensity, the return jumps and the variance jumps for the S\&P 500 Index and for the four models considered in this study. We use the S\&P 500 return series from 1990 to 2018.}
\label{fig:SP500_Results}
\end{sidewaysfigure}

Figure~\ref{fig:SP500_Results} shows the filtered annualized volatility (i.e., $\sqrt{V_t/252}$), the jump intensity, the return jumps and the variance jumps. Generally speaking, the volatility is rather similar between the various models, the only difference being that the SVCJ and SVCJSI models allow for sharper increases because of the volatility jumps. For instance, during the last financial crisis, the volatility goes higher with the models that permit variance jumps.

For models with constant jump intensity, the number of jumps per year is small---about three to five, on average. This figure is consistent with other studies such as \citet{eraker2003impact} and \citet{hurn2015estimating}. The filtered return jumps are rather similar for the SVYJ and SVCJ models.

On the other hand, the SVCJSI model allows for jump clustering as the jump intensity varies over time, and the jump processes are slightly different of those obtained with the SVYJ and SVCJ models. Specifically, the number of filtered jumps reduces during calm eras and increases during crises. 

\section{Concluding Remarks\label{sec:conclusion}}

In this paper, we developed a new estimation method for jump-diffusion models based upon the DNF method of \citet{kitagawa1987non}. The method we introduced allows for the likelihood evaluation of models that include stochastic volatility, return jumps, variance jumps, as well as stochastic jump intensity.

We then compared the performance of the DNF with the SIR. We found that the proposed DNF method is stable, accurate and faster than the SIR. We showed that a grid with 50 to 60 nodes leads to an MAPE of 0.1\%, on average, for the S\&P 500 index---a threshold that is reasonable. As an application of the method, we also found the maximum likelihood estimates of the SVCJSI model and its nested models with the S\&P 500 daily returns. The parameters obtained are consistent with those estimated in the literature.

Finally, this method is not only helpful for maximum likelihood estimation, but also for Bayesian inference as the likelihood function is an important building block when calculating posterior distributions. For instance, it would be interesting to combine the DNF with Markov chain Monte Carlo (MCMC) methods; this is left for future research.


\singlespacing
\begin{spacing}{0.5}
\bibliographystyle{rfs}
\bibliography{references}

\newcommand{\SortNoop}[1]{}
\begin{thebibliography}{48}
\providecommand{\natexlab}[1]{#1}
\providecommand{\url}[1]{\texttt{#1}}
\providecommand{\urlprefix}{URL }
\vspace{0.5cm}

\bibitem[{A{\"\i}t-Sahalia(2002)}]{ait2002maximum}
A{\"\i}t-Sahalia, Y. 2002.
\newblock Maximum likelihood estimation of discretely sampled diffusions: {A}
  closed-form approximation approach.
\newblock \emph{Econometrica} 70:223--262.

\bibitem[{A{\"\i}t-Sahalia and Kimmel(2007)}]{ai2007maximum}
A{\"\i}t-Sahalia, Y., and R.~Kimmel. 2007.
\newblock Maximum likelihood estimation of stochastic volatility models.
\newblock \emph{Journal of Financial Economics} 83:413--452.

\bibitem[{Andersen et~al.(2002)Andersen, Benzoni, and
  Lund}]{andersen2002empirical}
Andersen, T.~G., L.~Benzoni, and J.~Lund. 2002.
\newblock An empirical investigation of continuous-time equity return models.
\newblock \emph{Journal of Finance} 57:1239--1284.

\bibitem[{Andersen and S{\o}rensen(1996)}]{andersen1996gmm}
Andersen, T.~G., and B.~E. S{\o}rensen. 1996.
\newblock GMM estimation of a stochastic volatility model: A {M}onte {C}arlo
  study.
\newblock \emph{Journal of Business \& Economic Statistics} 14:328--352.

\bibitem[{Bakshi et~al.(1997)Bakshi, Cao, and Chen}]{bakshi1997empirical}
Bakshi, G., C.~Cao, and Z.~Chen. 1997.
\newblock Empirical performance of alternative option pricing models.
\newblock \emph{Journal of Finance} 52:2003--2049.

\bibitem[{Bardgett et~al.(2019)Bardgett, Gourier, and
  Leippold}]{bardgett2019inferring}
Bardgett, C., E.~Gourier, and M.~Leippold. 2019.
\newblock Inferring volatility dynamics and risk premia from the S\&P 500 and
  VIX markets.
\newblock \emph{Journal of Financial Economics} 131:593--618.

\bibitem[{Bartolucci and De~Luca(2001)}]{bartolucci2001maximum}
Bartolucci, F., and G.~De~Luca. 2001.
\newblock Maximum likelihood estimation of a latent variable time-series model.
\newblock \emph{Applied Stochastic Models in Business and Industry} 17:5--17.

\bibitem[{Bates(1996)}]{bates1996jumps}
Bates, D.~S. 1996.
\newblock Jumps and stochastic volatility: Exchange rate processes implicit in
  Deutsche Mark options.
\newblock \emph{Review of Financial Studies} 9:69--107.

\bibitem[{Bates(2000)}]{bates2000post}
Bates, D.~S. 2000.
\newblock Post-'87 crash fears in the {S\&P} 500 futures option market.
\newblock \emph{Journal of Econometrics} 94:181--238.

\bibitem[{Bates(2006)}]{bates2006maximum}
Bates, D.~S. 2006.
\newblock Maximum likelihood estimation of latent affine processes.
\newblock \emph{Review of Financial Studies} 19:909--965.

\bibitem[{B{\'e}gin et~al.(2019)B{\'e}gin, Dorion, and
  Gauthier}]{begin2019idiosyncratic}
B{\'e}gin, J.-F., C.~Dorion, and G.~Gauthier. 2019.
\newblock Idiosyncratic jump risk matters: Evidence from equity returns and
  options.
\newblock \emph{Forthcoming in the Review of Financial Studies\!} .

\bibitem[{Brandt and Santa-Clara(2002)}]{brandt2002simulated}
Brandt, M.~W., and P.~Santa-Clara. 2002.
\newblock Simulated likelihood estimation of diffusions with an application to
  exchange rate dynamics in incomplete markets.
\newblock \emph{Journal of Financial Economics} 63:161--210.

\bibitem[{Christoffersen et~al.(2010)Christoffersen, Jacobs, and
  Mimouni}]{christoffersen2010volatility}
Christoffersen, P., K.~Jacobs, and K.~Mimouni. 2010.
\newblock Volatility dynamics for the {S}\&{P} 500: {E}vidence from realized
  volatility, daily returns, and option prices.
\newblock \emph{Review of Financial Studies} 23:3141--3189.

\bibitem[{Christoffersen et~al.(2012)Christoffersen, Jacobs, and
  Ornthanalai}]{christoffersen2012dynamic}
Christoffersen, P., K.~Jacobs, and C.~Ornthanalai. 2012.
\newblock Dynamic jump intensities and risk premiums: Evidence from {S\&P} 500
  returns and options.
\newblock \emph{Journal of Financial Economics} 106:447--472.

\bibitem[{Clements et~al.(2004)Clements, Hurn, and
  White}]{clements2004discretised}
Clements, A., S.~Hurn, and S.~White. 2004.
\newblock Discretised non-linear filtering for dynamic latent variable models:
  With application to stochastic volatility.
\newblock \emph{Working Paper\!} .

\bibitem[{Clements et~al.(2006{\natexlab{a}})Clements, Hurn, White
  et~al.}]{clements2006estimating}
Clements, A., S.~Hurn, S.~White, et~al. 2006{\natexlab{a}}.
\newblock Estimating stochastic volatility models using a discrete non-linear
  filter.
\newblock \emph{Working Paper\!} .

\bibitem[{Clements et~al.(2006{\natexlab{b}})Clements, Hurn, and
  White}]{clements2006mixture}
Clements, A.~E., S.~Hurn, and S.~I. White. 2006{\natexlab{b}}.
\newblock Mixture distribution-based forecasting using stochastic volatility
  models.
\newblock \emph{Applied Stochastic Models in Business and Industry}
  22:547--557.

\bibitem[{Cont(2007)}]{cont2007volatility}
Cont, R. 2007.
\newblock Volatility clustering in financial markets: {E}mpirical facts and
  agent-based models.
\newblock In \emph{Long Memory in Economics}, pp. 289--309. Springer.

\bibitem[{Creal(2012)}]{creal2012survey}
Creal, D. 2012.
\newblock A survey of sequential Monte Carlo methods for economics and finance.
\newblock \emph{Econometric Reviews} 31:245--296.

\bibitem[{Danielsson(1994)}]{danielsson1994stochastic}
Danielsson, J. 1994.
\newblock Stochastic volatility in asset prices estimation with simulated
  maximum likelihood.
\newblock \emph{Journal of Econometrics} 64:375--400.

\bibitem[{Danielsson and Richard(1993)}]{danielsson1993accelerated}
Danielsson, J., and J.-F. Richard. 1993.
\newblock Accelerated {G}aussian importance sampler with application to dynamic
  latent variable models.
\newblock \emph{Journal of Applied Econometrics} 8:S153--S173.

\bibitem[{Duffie et~al.(2000)Duffie, Pan, and Singleton}]{duffie2000transform}
Duffie, D., J.~Pan, and K.~Singleton. 2000.
\newblock Transform analysis and asset pricing for affine jump-diffusions.
\newblock \emph{Econometrica} 68:1343--1376.

\bibitem[{Eraker(2001)}]{eraker2001mcmc}
Eraker, B. 2001.
\newblock {MCMC} analysis of diffusion models with application to finance.
\newblock \emph{Journal of Business \& Economic Statistics} 19:177--191.

\bibitem[{Eraker et~al.(2003)Eraker, Johannes, and Polson}]{eraker2003impact}
Eraker, B., M.~Johannes, and N.~Polson. 2003.
\newblock The impact of jumps in volatility and returns.
\newblock \emph{Journal of Finance} 58:1269--1300.

\bibitem[{Fridman and Harris(1998)}]{fridman1998maximum}
Fridman, M., and L.~Harris. 1998.
\newblock A maximum likelihood approach for non-{G}aussian stochastic
  volatility models.
\newblock \emph{Journal of Business \& Economic Statistics} 16:284--291.

\bibitem[{Gordon et~al.(1993)Gordon, Salmond, and Smith}]{gordon1993novel}
Gordon, N.~J., D.~J. Salmond, and A.~F. Smith. 1993.
\newblock Novel approach to nonlinear/non-Gaussian Bayesian state estimation.
\newblock In \emph{IEEE Proceedings F (Radar and Signal Processing)}, vol. 140,
  pp. 107--113.

\bibitem[{Harvey et~al.(1994)Harvey, Ruiz, and
  Shephard}]{harvey1994multivariate}
Harvey, A., E.~Ruiz, and N.~Shephard. 1994.
\newblock Multivariate stochastic variance models.
\newblock \emph{Review of Economic Studies} 61:247--264.

\bibitem[{Heston(1993)}]{heston1993}
Heston, S. 1993.
\newblock A closed-form solution for options with stochastic volatility with
  applications to bond and currency options.
\newblock \emph{Review of Financial Studies} 6:327.

\bibitem[{Hurn et~al.(2015)Hurn, Lindsay, and McClelland}]{hurn2015estimating}
Hurn, A.~S., K.~A. Lindsay, and A.~J. McClelland. 2015.
\newblock Estimating the parameters of stochastic volatility models using
  option price data.
\newblock \emph{Journal of Business \& Economic Statistics} 33:579--594.

\bibitem[{Jacquier et~al.(1994)Jacquier, Polson, and
  Rossi}]{jacquier1994bayesian}
Jacquier, E., N.~G. Polson, and P.~E. Rossi. 1994.
\newblock Bayesian analysis of stochastic volatility models.
\newblock \emph{Journal of Business \& Economic Statistics} 12:371--389.

\bibitem[{Johannes et~al.(1999)Johannes, Kumar, and Polson}]{johannes1999state}
Johannes, M., R.~Kumar, and N.~G. Polson. 1999.
\newblock State dependent jump models: {H}ow do US equity indices jump.
\newblock \emph{Woking paper\!} .

\bibitem[{Johannes et~al.(2009)Johannes, Polson, and
  Stroud}]{johannes2009optimal}
Johannes, M.~S., N.~G. Polson, and J.~R. Stroud. 2009.
\newblock Optimal filtering of jump diffusions: Extracting latent states from
  asset prices.
\newblock \emph{Review of Financial Studies} 22:2759--2799.

\bibitem[{Kalman(1960)}]{kalman1960new}
Kalman, R.~E. 1960.
\newblock A new approach to linear filtering and prediction problems.
\newblock \emph{Journal of Basic Engineering} 82:35--45.

\bibitem[{Kitagawa(1987)}]{kitagawa1987non}
Kitagawa, G. 1987.
\newblock Non-{G}aussian state-space modeling of nonstationary time series.
\newblock \emph{Journal of the American Statistical Association} 82:1032--1041.

\bibitem[{Langrock et~al.(2012)Langrock, MacDonald, and
  Zucchini}]{langrock2012some}
Langrock, R., I.~L. MacDonald, and W.~Zucchini. 2012.
\newblock Some nonstandard stochastic volatility models and their estimation
  using structured hidden {M}arkov models.
\newblock \emph{Journal of Empirical Finance} 19:147--161.

\bibitem[{Maheu and McCurdy(2004)}]{maheu2004news}
Maheu, J.~M., and T.~H. McCurdy. 2004.
\newblock News arrival, jump dynamics, and volatility components for individual
  stock returns.
\newblock \emph{Journal of Finance} 59:755--793.

\bibitem[{Malik and Pitt(2011)}]{malik2011particle}
Malik, S., and M.~K. Pitt. 2011.
\newblock Particle filters for continuous likelihood evaluation and
  maximisation.
\newblock \emph{Journal of Econometrics} 165:190--209.

\bibitem[{Melino and Turnbull(1990)}]{melino1990pricing}
Melino, A., and S.~M. Turnbull. 1990.
\newblock Pricing foreign currency options with stochastic volatility.
\newblock \emph{Journal of Econometrics} 45:239--265.

\bibitem[{Merton(1976)}]{merton1976option}
Merton, R.~C. 1976.
\newblock Option pricing when underlying stock returns are discontinuous.
\newblock \emph{Journal of Financial Economics} 3:125--144.

\bibitem[{Nelson(1988)}]{nelson1988time}
Nelson, D.~B. 1988.
\newblock \emph{The time series behavior of stock market volatility and
  returns}.
\newblock Ph.D. thesis, Massachusetts Institute of Technology.

\bibitem[{Ornthanalai(2014)}]{ornthanalai2014levy}
Ornthanalai, C. 2014.
\newblock Levy jump risk: Evidence from options and returns.
\newblock \emph{Journal of Financial Economics} 112:69--90.

\bibitem[{Pan(2002)}]{pan2002jump}
Pan, J. 2002.
\newblock The jump-risk premia implicit in options: Evidence from an integrated
  time-series study.
\newblock \emph{Journal of Financial Economics} 63:3--50.

\bibitem[{Pitt et~al.(2014)Pitt, Malik, and Doucet}]{pitt2014simulated}
Pitt, M.~K., S.~Malik, and A.~Doucet. 2014.
\newblock Simulated likelihood inference for stochastic volatility models using
  continuous particle filtering.
\newblock \emph{Annals of the Institute of Statistical Mathematics}
  66:527--552.

\bibitem[{Rémillard(2013)}]{remillard2013statistical}
Rémillard, B. 2013.
\newblock \emph{Statistical methods for financial engineering}.
\newblock Chapman \& Hall.

\bibitem[{Shephard(1993)}]{shephard1993fitting}
Shephard, N. 1993.
\newblock Fitting nonlinear time-series models with applications to stochastic
  variance models.
\newblock \emph{Journal of Applied Econometrics} 8:S135--S152.

\bibitem[{Taylor(1986)}]{taylor2008modelling}
Taylor, S.~J. 1986.
\newblock \emph{Modelling financial time series}.
\newblock World Scientific.

\bibitem[{Todorov and Tauchen(2011)}]{todorov2011volatility}
Todorov, V., and G.~Tauchen. 2011.
\newblock Volatility jumps.
\newblock \emph{Journal of Business \& Economic Statistics} 29:356--371.

\bibitem[{Watanabe(1999)}]{watanabe1999non}
Watanabe, T. 1999.
\newblock A non-linear filtering approach to stochastic volatility models with
  an application to daily stock returns.
\newblock \emph{Journal of Applied Econometrics} 14:101--121.

\end{thebibliography}
\end{spacing}
\doublespacing

\appendix
\setcounter{section}{0}
\justifying


\clearpage

\newpage
\setcounter{page}{1}

\center{\Large \bf  Supplementary Material}
\singlespacing

\setcounter{section}{0}
\justifying
\setlength{\parindent}{0pt}
\setlength{\parskip}{0.5\baselineskip}

\setcounter{page}{1}
\setcounter{section}{0}
\setcounter{equation}{0}
\renewcommand{\theequation}{SM.\arabic{equation}}
\renewcommand*{\thepage}{SM-\arabic{page}}

\setcounter{figure}{0}
\makeatletter
\renewcommand{\thefigure}{SM.\@arabic\c@figure}
\makeatother

\setcounter{table}{0}
\makeatletter
\renewcommand{\thetable}{SM.\@arabic\c@table}
\makeatother

\setcounter{footnote}{0}

\section{Particle Filter\label{secsm:pf}}

In this section, we briefly describe the sequential Monte Carlo method used to compute the likelihood function in this study. This implementation of the particle filter closely follows the work of \citet{gordon1993novel}; specifically, we use the bootstrap filter. Algorithm 2 describes the various steps of the general SIR method.
\vspace{0.2cm}

\begin{algorithm}
\small
\caption{Sequential Importance Resampling}
\begin{algorithmic}[1]
  \STATE initiate the state, $\mathbf{x}_0^{(i)} \sim p\left( \, \cdot \, \right)$ for $i \in \{1,2,...,N_p\}$, where $p$ is the initial state density
  \FOR{each $t \in \{1, ..., T\}$}
  \FOR{each $i \in \{1, ..., N_p\}$}
  \STATE sample $\tilde{\mathbf{x}}_t^{(i)} \sim q\left(\,  \cdot\,\, \middle|\, \mathbf{x}_{t-1}^{(i)}  \right)$
  \STATE set $\tilde{\mathbf{x}}_{0:t}^{(i)} = \left( {\mathbf{x}}_{0:t-1}^{(i)}, \tilde{\mathbf{x}}_{t}^{(i)} \right)$
  \STATE evaluate the importance weights, i.e., $w_t \left( \tilde{\mathbf{x}}_{0:t}^{(i)} \right) = w_{t-1}^{(i)} \, r\left( \mathbf{y}_t \, \middle|\, \tilde{\mathbf{x}}_{t}^{(i)} \right)$
  \STATE normalize the importance weights, i.e., $\widehat{w}_t^{(i)} = \frac{w_t \left( \tilde{\mathbf{x}}_{0:t}^{(i)} \right)}{\sum_{j=1}^N w_t \left( \tilde{\mathbf{x}}_{0:t}^{(j)} \right)}$
  \ENDFOR
  \STATE resample with replacement $N_p$ particles from $\tilde{\mathbf{x}}_{0:t}^{(i)}$ according to the weights $\widehat{w}_t^{(i)}$
  \STATE set $w_t^{(i)} = \frac{1}{N}$ for each $i \in \{1,...,N\}$
  \STATE compute the time-$t$ likelihood contribution, i.e., $f\left(\mathbf{y}_t \, \middle|\, \mathbf{y}_{1:t-1} \right) \approx \sum_{i=1}^N w_t \left( \tilde{\mathbf{x}}_{0:t}^{(i)} \right)$
  \ENDFOR
  \STATE compute the likelihood function $\mathcal{L}(\Theta)$ by taking the product of the likelihood contributions
\end{algorithmic}
\end{algorithm}
\vspace{-0.1cm}
\normalsize

\section{Study of the DNF Bias\label{secsm:bias}}

The efficacy of the DNF method for estimating the parameters of jump-diffusion models using maximum likelihood estimation is now assessed. In this experiment, we compare the parameter estimates obtained by using DNF with their respective \emph{true} values. The primary objective of this simulation exercise is to assess whether the DNF leads to systematic biases in the estimation parameters.

Specifically, for a set of likely parameters---consistent with most studies cited in the introduction of this study---we generate 100 ten-year return paths, i.e., $T = 2,520$. Then, we use the DNF method coupled with a numerical optimizer to find the maximum likelihood estimates.

Table~\ref{tab:bias} reports the average value, the bias as well as the root mean square error for each parameter and each model. In summary, the average is very close to the true value for most of the parameters, leading to small biases. The only exception is parameter $\kappa$: the bias associated with this parameter hovers between 0.25 and 0.75. Yet, even though this bias looks large at first sight, it is not so dreadful. In fact, when expressed in terms of daily persistence of the process---as it is typically done in econometrics---the difference is rather small, e.g., $e^{-3.717 h} = 0.9882$ versus $e^{-3.000 h} = 0.9854$ for the SV model.

\begin{table}
\footnotesize
\topcaption{\textbf{Study of the Bias per Model.}}
{\centering
\begin{tabularx}{\linewidth}{lCCCC}
\toprule
\multicolumn{5}{l}{\textbf{Panel A: SV Model.}} \\
\midrule
      & \textbf{Average} & \textbf{True Value} & \textbf{Bias} & \textbf{RMSE} \\
\cmidrule(lr){2-2} \cmidrule(lr){3-3} \cmidrule(lr){4-4} \cmidrule(lr){5-5}
$\mu$ & $\hphantom{-}$0.033 & $\hphantom{-}$0.060 & $-0.027$ & 0.042 \\[-0.2ex]
$\kappa$ & $\hphantom{-}$3.717 & $\hphantom{-}$3.000 & $\hphantom{-}$0.717 & 1.117 \\[-0.2ex]
$\theta$ & $\hphantom{-}$0.033 & $\hphantom{-}$0.030 & $\hphantom{-}$0.003 & 0.005 \\[-0.2ex]
$\sigma$ & $\hphantom{-}$0.300 & $\hphantom{-}$0.300 & $\hphantom{-}$0.000 & 0.031 \\[-0.2ex]
$\rho_v$ & $-$0.599 & $-0.600$ & $\hphantom{-}$0.001 & 0.070 \\[0.2ex]
\toprule
\multicolumn{5}{l}{\textbf{Panel B: SVYJ Model.}} \\
\midrule
      & \textbf{Average} & \textbf{True Value} & \textbf{Bias} & \textbf{RMSE} \\
\cmidrule(lr){2-2} \cmidrule(lr){3-3} \cmidrule(lr){4-4} \cmidrule(lr){5-5}
$\mu$ & $\hphantom{-}$0.040    & $\hphantom{-}$0.060 & $-0.020$ & 0.045 \\[-0.2ex]
$\kappa$ & $\hphantom{-}$3.693 & $\hphantom{-}$3.000 & $\hphantom{-}$0.693 & 1.116 \\[-0.2ex]
$\theta$ & $\hphantom{-}$0.033 & $\hphantom{-}$0.030 & $\hphantom{-}$0.003 & 0.005 \\[-0.2ex]
$\sigma$ & $\hphantom{-}$0.300 & $\hphantom{-}$0.300 & $\hphantom{-}$0.000 & 0.033 \\[-0.2ex]
$\rho_v$ & $-0.599 $           & $-0.600$ & $\hphantom{-}$0.001 & 0.081 \\[-0.2ex]
$\theta$ & $\hphantom{-}$4.909 & $\hphantom{-}$5.000 & $-0.091$ & 1.780 \\[-0.2ex]
$\alpha$ & $-0.022$            & $-0.020$ & $-0.002$ & 0.009 \\[-0.2ex]
$\delta$ & $\hphantom{-}$0.030             & $\hphantom{-}$0.030 & $\hphantom{-}$0.000 & 0.005 \\[0.2ex]
\toprule
\multicolumn{5}{l}{\textbf{Panel C: SVCJ Model.}} \\
\midrule
      & \textbf{Average} & \textbf{True Value} & \textbf{Bias} & \textbf{RMSE} \\
\cmidrule(lr){2-2} \cmidrule(lr){3-3} \cmidrule(lr){4-4} \cmidrule(lr){5-5}
$\mu$ & $\hphantom{-}$0.059 & $\hphantom{-}$0.060 & $-$0.001 & 0.038 \\[-0.2ex]
$\kappa$ & $\hphantom{-}$3.530 & $\hphantom{-}$3.000 & $\hphantom{-}$0.530 & 0.778 \\[-0.2ex]
$\theta$ & $\hphantom{-}$0.031 & $\hphantom{-}$0.030 & $\hphantom{-}$0.001 & 0.008 \\[-0.2ex]
$\sigma$ & $\hphantom{-}$0.305 & $\hphantom{-}$0.300 & $\hphantom{-}$0.005 & 0.043 \\[-0.2ex]
$\rho_v$ & $-$0.598 & $-$0.600 & $\hphantom{-}$0.002 & 0.105 \\[-0.2ex]
$\omega$ & $\hphantom{-}$5.122 & $\hphantom{-}$5.000 & $\hphantom{-}$0.122 & 1.791 \\[-0.2ex]
$\alpha$ & $-$0.024 & $-$0.020 & $-$0.004 & 0.012 \\[-0.2ex]
$\delta$ & $\hphantom{-}$0.026 & $\hphantom{-}$0.030 & $-$0.004 & 0.007 \\[-0.2ex]
$\nu$ & $\hphantom{-}$0.011 & $\hphantom{-}$0.010 & $\hphantom{-}$0.001 & 0.005 \\[-0.2ex]
$\rho_z$ & $-$0.887 & $-$1.000 & $\hphantom{-}$0.113 & 0.591 \\[0.2ex]
\toprule
\multicolumn{5}{l}{\textbf{Panel D: SVCJSI Model.}} \\
\midrule
      & \textbf{Average} & \textbf{True Value} & \textbf{Bias} & \textbf{RMSE} \\
\cmidrule(lr){2-2} \cmidrule(lr){3-3} \cmidrule(lr){4-4} \cmidrule(lr){5-5}
 $\mu$   & $\hphantom{-}$0.058 & $\hphantom{-}$0.060 & $-$0.002 & 0.010 \\[-0.2ex]
$\kappa$ & $\hphantom{-}$3.261 & $\hphantom{-}$3.000 & $\hphantom{-}$0.261 & 0.451 \\[-0.2ex]
$\theta$ & $\hphantom{-}$0.031 & $\hphantom{-}$0.030 & $\hphantom{-}$0.001 & 0.005 \\[-0.2ex]
$\sigma$ & $\hphantom{-}$0.276 & $\hphantom{-}$0.300 & $-$0.024 & 0.033 \\[-0.2ex]
$\rho_v$ & $-$0.577              & $-$0.600 & $\hphantom{-}$0.023 & 0.091 \\[-0.2ex]
$\chi$   & $\hphantom{-}$3.123 & $\hphantom{-}$3.000 & $\hphantom{-}$0.123 & 0.373 \\[-0.2ex]
$\omega$ & $\hphantom{-}$5.060 & $\hphantom{-}$5.000 & $\hphantom{-}$0.060 & 0.780 \\[-0.2ex]
$\xi$    & $\hphantom{-}$5.089 & $\hphantom{-}$5.000 & $\hphantom{-}$0.089 & 0.817 \\[-0.2ex]
$\rho_l$ & $-$0.309              & $-$0.300 & $-$0.009 & 0.037 \\[-0.2ex]
$\alpha$ & $-$0.020              & $-$0.020 & $\hphantom{-}$0.000 & 0.003 \\[-0.2ex]
$\delta$ & $\hphantom{-}$0.028 & $\hphantom{-}$0.030 & $-$0.002 & 0.005 \\[-0.2ex]
$\nu$    & $\hphantom{-}$0.010 & $\hphantom{-}$0.010 & $\hphantom{-}$0.000 & 0.002 \\[-0.2ex]
$\rho_z$ & $-$1.004              & $-$1.000 & $-$0.004 & 0.130 \\
\bottomrule
\end{tabularx}}
\label{tab:bias}
\footnotesize

This table reports the average parameter, the true value, the bias as well as the root mean square error, denoted by RMSE. For a set of likely parameters, we generate 100 ten-year return paths, i.e., $T = 2,520$. Then, we use the DNF method coupled with a numerical optimizer to find the maximum likelihood estimates.
\end{table}
\normalsize

\end{document}